\pdfoutput=1
\documentclass[prb,aps,floatfix,twocolumn]{revtex4}%
\usepackage{amsfonts}
\usepackage{amsmath}
\usepackage{amssymb}
\usepackage{graphicx}%
\begin{document}
\preprint{HEP/123-qed}
\title[Mixed valency in cerium oxide crystallographic phases: Determination of
valence of the different cerium sites by the bond valence method]{Mixed valency in cerium oxide crystallographic phases: Determination of
valence of the different cerium sites by the bond valence method}
\author{E. Shoko, M. F. Smith, and Ross H. McKenzie}
\affiliation{University of Queensland, Department of Physics, Brisbane, QLD 4072, Australia}
\pacs{PACS number}

\begin{abstract}
We have applied the bond valence method to cerium oxides to determine the
oxidation states of the Ce ion at the various site symmetries of the crystals.
The crystals studied include cerium dioxide and the two sesquioxides along
with some selected intermediate phases which are crystallographically well
characterized. Our results indicate that cerium dioxide has a mixed-valence
ground state with an $f$-electron population on the Ce site of $0.27$ while
both the $A$- and $C$-sesquioxides have a nearly pure ${f^{1}}$ configuration.
The Ce sites in most of the intermediate oxides have non-integral valences.
Furthermore, many of these valences are different from the values predicted
from a naive consideration of the stoichiometric valence of the compound.

\end{abstract}
\volumeyear{year}
\volumenumber{number}
\issuenumber{number}
\eid{identifier}
\date[Date text]{date}
\received[Received text]{date}

\revised[Revised text]{date}

\accepted[Accepted text]{date}

\published[Published text]{date}

\startpage{01}
\endpage{102}
\maketitle


\section{Introduction}

\label{sec:intro}

Mixed valency of transition metal and rare-earth ions in solids and compounds
is a question of fundamental interest in materials physics, chemisty, and
molecular biophysics. \cite{Clark2008a} In 1967, Robin and Day \cite{Robin1967} published a classification scheme for
mixed-valence that is still widely used today.\cite{Cox1987} Class 1 describes
systems with two crystallographic sites that are clearly distinct and and the
two sites have integral but unequal valence. There is a large energy
associated with transfer of electrons between sites. At the other extreme is
Class 3 for which there are two sites which are not distinguishable, and one
assigns a non-integral valence to both sites. The valence electrons are
delocalised between the two sites. Class 2 is the intermediate case where the
environments of the two sites are distinguishable but not very different. The
energy associated with electron transfer is sufficiently small that it can be
thermally activated and be associated with significant optical absorption in
the visible range. On the time scale of the vibrations of the atoms the
electrons may appear to be delocalised. Classes 1 and 2 correspond to what Varma \cite{Varma1976} terms
\emph{inhomogeneous} mixed valence, although, perhaps, \emph{inhomogeneous integral} valence may be more appropriate. Class 3 corresponds to \emph{homogeneous}
mixed valence. 

In systems which are characterized by homogeneous mixed-valence, each ion has
the same, noninteger, valence which is a result of a quantum mechanical
superposition of two integral valences occuring on each ion (See for example
Eq. (\ref{valence}). Compounds exhibiting this type of mixed-valence
include, for example, CePd$_{3}$ (Ref. \onlinecite{Gardner1972}), TmSe (Ref.
\onlinecite{Batlogg1979}) and SmB$_{6}$ (Ref. \onlinecite{Adroja1991}) where
the valences of the ions are $3.45$, $2.72$ and $3.7$ for Ce, Tm, and Sm
respectively. In TmSe, the valence of $2.72$ for the Tm ion is a result of
valence fluctuations of this ion between the Tm$^{2+}$ and Tm$^{3+}$ states.
\cite{Batlogg1979} In contrast, the inhomogeneous mixed-valence case involves
a mixture of different integer valence ions which occupy inequivalent lattice
sites in a static charge-ordered array. Examples of this are provided by
Fe$_{3}$O$_{4}$ (Ref. \onlinecite{Walz2002}), Eu$_{3}$O$_{4}$ (Ref.
\onlinecite{Batlogg1975}) and Eu$_{3}$S$_{4}$ (Ref. \onlinecite{Ohara2004}).
The inverse spinel crystal structure of magnetite (Fe$_{3}$O$_{4}$) is considered the classic case of inhomogeneous
mixed-valence. In this crystal, the Fe$^{3+}$ ions completely occupy the
tetrahedral ($A$) sublattice while the octahedral ($B$) sublattice is equally
shared between the Fe$^{3+}$ and Fe$^{2+}$ ions and the ionic formulation is
$\left(  \text{Fe}^{2+}\right)  \left(  \text{Fe}^{3+}\right)  _{2}\left(
\text{O}^{2-}\right)  _{4}$. However, we note that this inverse-spinel charge
ordering in Fe$_{3}$O$_{4}$ has recently been challenged in favour of the
normal spinel charge structure where the Fe$^{3+}$ ions exclusively occupy all
the octahedral sites while the Fe$^{2+}$ ions reside in the tetrahedral sites.
\cite{Ravindran2008} The crystal of Eu$_{3}$O$_{4}$ is a good example of the
case where ions of different valence strictly occupy inequivalent cation
sites. It consists of two nonequivalent Eu sites in which the Eu$^{2+}$ and
Eu$^{3+}$ ions occupy the eight- and six-coordinated sites respectively giving
a static charge-ordered array whose ionic formulation may be written as
$\left(  \text{Eu}^{2+}\right)  \left(  \text{Eu}^{3+}\right)  _{2}\left(
\text{O}^{2-}\right)  _{4}$ (Ref. \onlinecite{Batlogg1975}).

Oxides of cerium (Ce) appear to exhibit mixed-valence characteristics which
may help to explain some of their properties relevant to their engineering
applications. An important industrial use of Ce oxides is as anode materials
in high-temperature solid-oxide fuel cells. \cite{Trovarelli2002a} For these
applications, CeO$_{2}$ (ceria) and Ce$_{2}$O$_{3}$ represent the extremal
oxidation states in the reversible chemical reaction, Eq. (\ref{reduce}):%

\begin{equation}
\text{CeO}_{2}\rightleftharpoons\text{CeO}_{2-y}+\frac{y}{2}\text{O}%
_{2}\text{, \ \ \ \ }0\leq y\leq0.5 \label{reduce}%
\end{equation}
Strong electron-correlation effects may be involved in the crystallographic
phase transitions which occur between CeO$_{2}$ and Ce$_{2}$O$_{3}$, i.e.,
the reaction in Eq. (\ref{reduce}). When CeO$_{2}$ is reduced to the various
defective phases, CeO$_{2-y}$, O vacancies are formed in the lattice
structure. The crystal structure adopted by any such defective phase,
CeO$_{2-y}$, is understood to be the one that provides the most favourable
energetics for the arrangement of all the O vacancies within the structure. In
a widely accepted view of the microscopic description of O vacancy formation
and ordering in CeO$_{2-y}$ phases, the two electrons left by the O atom when
an O vacancy forms fully localize on two of the nearest Ce$^{4+}$ ions. \cite{Hoskins1995,
Skorodumova2002, Trovarelli2002a, Esch2005} The localization of an electron on
a Ce$^{4+}$ ion converts it to the slightly larger Ce$^{3+}$ ion with one
electron in the $4f$ orbital. In the reverse process where a defective phase,
CeO$_{2-y}$, is oxidized, two $4f$ electrons are transferred from the two
neighbouring Ce$^{3+}$ ion sites into the O ${2p}$ band.

This description leads one to expect that the Ce lattice sites in the
defective CeO$_{2-y}$ phases would consist of a mixture of Ce$^{3+}$ and
Ce$^{4+}$ ions in a static charge-ordered array. Thus, useful insight into the
microscopic processes involved in the reversible chemical reaction,
(\ref{reduce}), could be gained from a knowledge of the valences of the Ce
ions in the defective CeO$_{2-y}$ phases. However, as we discuss in the next
section, it turns out that the task of establishing the oxidation states of
the Ce ions in the lattices of the crystal phases involved in (\ref{reduce})
is very challenging owing to valence fluctuations on the Ce ions.

Here, we report the results of calculations based on a simple empirical
method, the bond valence model \cite{Brown1992,Brown2002,Burdett1995a}, to
determine the $f$-electron occupancies in the various Ce ion sites for seven
of the crystallographic phases involved in Eq. (\ref{reduce}). The
crystals studied include $A$-Ce$_{2}$O$_{3}$, $C$-Ce$_{2}$O$_{3}$, Ce$_{3}%
$O$_{5}$, Ce$_{7}$O$_{12}$, Ce$_{11}$O$_{20}$, Ce$_{6}$O$_{11}$ and CeO$_{2}$.

The structure of this paper is as follows. In Section \ref{sec:MVProblem}, we
outline the general problem of mixed-valence in Ce oxides from both
experimental and theoretical perspectives. Section \ref{sec:BVM} describes the
bond valence model and how we used it to determine the valences of Ce ions in
the different phases of the Ce oxide crystals. We present our results in
Section \ref{sec:results} which we then discuss in light of two other models
for predicting cationic valences in crystals. The main conclusions of this
paper are given in Section \ref{sec:conclusion}. Appendix A lists all the Ce-O
bond lengths used in our calculations for Ce-centred polyhedra.

\section{The Mixed-Valence Problem in Cerium Oxides}

\label{sec:MVProblem}

In Ce oxides, the number of $f$ electrons on a Ce site, $N_{f}$, is observed
to lie between $0$ and slightly above $1.0$. An isolated Ce atom has the
electronic configuration [Xe]$4f^{1}5d^{1}6s^{2}$. In a crystal of the oxide,
the $5d^{1}6s^{2}$ electrons participate in bonding, forming part of the O $2p$
valence band, while the state of the $4f^{1}$ electron depends on the energy
of the associated electronic state within the particular crystal structure. If
the $4f^{1}$ state lies well within the band gap between the empty $5d6s$
conduction band and the O $2p$ valence band, then the $4f^{1}$ electron is
fully localized on the Ce atom and $N_{f}=1.0$. On the other hand, if the
$4f^{1}$ electron state merges with the valence band, then the electron is
lost to neighbouring O atoms so that $N_{f}=0$ thus forming an ionic bond.
Intermediate behaviour with fractional occupancy of the $4f$ state occurs when
the associated state lies close to the valence band. Occupancies for which
$N_{f}>1.0$ are however energetically discouraged by the large on-site Coulomb
repulsion for localized Ce states.

\begin{figure}
[ptb]
\begin{center}
\includegraphics[natheight=5.333300in,
natwidth=6.106400in,
height=2.6948in,
width=3.0805in]
{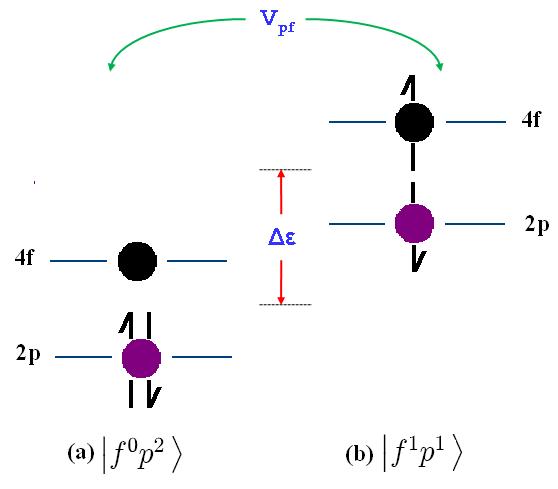}%
\caption{A schematic of the energy level structure of CeO$_{2}$ showing the
two important system parameters in the Anderson impurity model: $\Delta
\varepsilon$, the energy gap between the two states ${\left\vert f^{0}p^{2}\right\rangle}$ and ${\left\vert f^{1}p^{1}\right\rangle}$ which are mixed by the hybridization, $V_{pf}$. For CeO$_{2}$, the two parameters, $\Delta
\varepsilon$ and $V_{pf}$ are comparable and the Ce $4f$ level may be close to
the O $2p$ level. \cite{Kotani1988} In (a), the pure $f^{0}$ configuration
corresponding to Ce$^{4+}$ at the Ce site is shown. In (b), an electron has hopped from the O $2p$ site to the Ce site where it occupies the $4f$ level giving the pure $f^{1}$ configuration, i.e., Ce$^{3+}$ as shown. An electron hole is thus created in the O $2p$ valence band. The mixing between these states due to the hybridization, $V_{pf}$, may result in mixed-valence at the Ce site. Here, we have set the on-site Coulomb
repulsion, $U_{ff}$, (for double occupancy of the $f$ orbital) to infinity. }%
\label{hybrid}%
\end{center}
\end{figure}

Fig. \ref{hybrid} illustrates this situation for CeO$_{2}$ showing the
relevant energy scales when the system is described by the Anderson impurity
model. \cite{Kotani1988} Here, $U_{ff}$, the on-site Coulomb repulsion for the
$f$-orbital on a Ce site is considered to be very large. As already noted
above, setting $U_{ff}\rightarrow\infty$ excludes from the ground state
wavefunction the state $\left\vert f^{2}p^{0}\right\rangle $ which corresponds
to $N_{f}=2.0$. Once this assumption has been made, then, as shown in Fig.
\ref{hybrid}, the only key parameters of the model become the energy gap
between the $\left\vert f^{1}p^{1}\right\rangle $ and $\left\vert f^{0}%
p^{2}\right\rangle $ configurations, $\Delta\varepsilon$, and the
hybridization strength, $V_{pf}$.%

Due to the mixing of states shown in Fig. \ref{hybrid}, the electronic ground
state wavefunction of a Ce site in the oxide can be written in the general form:%

\begin{equation}
\Psi=\alpha\left\vert f^{0}p^{2}\right\rangle +\beta\left\vert f^{1}%
p^{1}\right\rangle \label{valence}%
\end{equation}
where $\alpha$ and $\beta$ are constants and $\alpha^{2}+\beta^{2}=1$, with
the states assumed to be orthogonal. If for the two states $\left\vert f^{0}%
p^{2}\right\rangle $ and $\left\vert f^{1}p^{1}\right\rangle $, $\Delta
\varepsilon$ and $V_{pf}$ are comparable, then the necessary conditions for valence fluctuation phenomena
hold. \cite{Varma1976,Lawrence1981} Thus one can think of the $4f$-level at a
given Ce site as fluctuating between the $f^{0}$ and $f^{1}$ configurations.

Various experimental and theoretical approaches have been brought to bear on
the problem of mixed valence in Ce oxides with much of the focus concentrated
on CeO$_{2}$ and to some extent Ce$_{2}$O$_{3}$. There appears to be general
agreement that the electronic ground state of Ce$_{2}$O$_{3}$ is the $f^{1}$
configuration. \cite{Kotani1992,Groot2008} However, controversy exists on the
exact nature of the ground state of CeO$_{2}$ with strong arguments in favour
of both a pure $f^{0}$ configuration and a mixed-valence ground state. We
briefly review some of this interesting debate indicating which approaches
have consistently arrived at the same conclusions and which ones have had
mixed interpretations.

The $3d$ photoabsorption and photoemission spectra analysed by the Anderson
impurity model consistently reach the conclusion that CeO$_{2}$ has a mixed
valent ground state.
\cite{Wuilloud1984,Jo1985,Kotani1985,Kotani1988,Kotani1992,Groot2008,Nakazawa1996}
A cluster model has also been used to interpret some $3d$ photoemission
spectra and it was also concluded that the $f$ state is strongly mixed-valent
in the ground state of CeO$_{2}$. \cite{Fujimori1983,Fujimori1983a} Although
the results of $3d$ x-ray photoemission spectroscopy (XPS) interpreted in the
Anderson impurity model have consistently predicted a mixed-valent ground
state for CeO$_{2}$, doubts have been raised about the reliability of
assigning initial state configurations from Ce $3d$ XPS spectra because of the
possibility of reduction of the oxide on exposure to the x-ray radiation.
\cite{Rao1997} Perhaps, an even more difficult challenge in the interpretation
of XPS spectra is how to trace the ground state configuration of a sample from
the spectral signatures of the final state of XPS. The conventional
interpretation is that in Ce$_{2}$O$_{3}$, the two-peak structure observed in
XPS spectra is a result of final state effects whereas CeO$_{2}$ shows a
three-peak structure in its XPS spectrum which has been attributed to final
state effects. \cite{Kotani1988,Kotani1992,Groot2008}

X-ray absorption near-edge spectroscopy (XANES) has provided a second source
of evidence for the mixed valence of the ground state of CeO$_{2}$. The
characteristic double peak structure observed in all XANES spectra is
considered a signature for mixed valence in CeO$_{2}$. \cite{Soldatov1994,
Dexpert1987,Bianconi1987,Karnatak1987,Bianconi1982,Finkelstein1992,Krill1981,Kotani1987}%

In contrast to the $3d$ XPS and XANES data discussed above, the $4d$-$4f$
photoabsorption spectra of CeO$_{2}$ support the $f^{0}$ ground state
configuration based on their close resemblance to La trihalide spectra which
have no $f$ electron while they differ considerably from the Ce trihalide
spectra which have one $f$ electron. \cite{Haensel1970,Hanyuu1985} Also in
favour of the $f^{0}$ ground state configuration are results from reflectance
spectroscopy in the $4d$-$4f$ absorption region. \cite{Miyahara1987}

Among the methods which have produced an ambiguous picture of the ground state
of CeO$_{2}$ are included resonant inelastic x-ray spectroscopy (RIXS),
bremsstrahlung isochromat spectroscopy (BIS), valence band XPS and energy band
structure calculations. Application of resonant photoemission to the problem
led to the conclusion that the ground state of CeO$_{2}$ is mixed-valent by
some authors \cite{Matsumoto1994,Butorin1996} while a pure $f^{0}$
configuration was claimed by others. \cite{Sham2005a,Haensel1970,Hanyuu1985}
The spectrum investigated by the bremstrahlung isochromat spectroscopy (BIS)
combined with XPS study shows empty localized $4f$ states in the band gap
providing evidence in support of the pure $f^{0}$ configuration.
\cite{Wuilloud1984,Allen1985} This conclusion was also supported by valence
band XPS studies. \cite{Orchard1977,Ryzhkov1985} However, it was later shown
that the valence-band photoemission, BIS, and $4d$ photoabsorption spectra can
all be explained consistently with other core-level spectra by the Anderson
impurity model. \cite{Nakano1987a,Jo1988,Nakazawa1996} The results of this
analysis led to the conclusion that CeO$_{2}$ is mixed-valent in the ground state.

An energy band calculation by the linear augmented plane wave (LAPW) method
showed a $4f$ electron count of $\sim0.5$ with considerable covalent
character. \cite{Koelling1983} Similar results were obtained from a linear
muffin-tin orbital (LMTO) band structure \cite{Finkelstein1992} and in a
molecular orbital calculation of a CeO$_{8}$ cluster. \cite{Thornton1981} On
the other hand, band structures of CeO$_{2}$ obtained from density functional
theory (DFT) calculations, both in the local density approximation (LDA) and
generalized gradient approximation (GGA), have empty $4f$ states above the
valence band supporting the $f^{0}$ ground state configuration.
\cite{Hay2006,Andersson2007,Fabris2005a,Skorodumova2001} This electronic
configuration for CeO$_{2}$ is widely accepted in DFT work largely because it
predicts the structural properties of the CeO$_{2}$ crystal with reasonable
accuracy. However, LDA and GGA do not give the experimentally observed band
structure and structural properties of Ce$_{2}$O$_{3}$. Some pragmatic
strategies have been adopted to augment LDA and GGA so that predictions closer
to the experimental results could be obtained. Skorodumova \textit{et al}.
\cite{Skorodumova2001} artificially localized the $4f$ states in what was
called the `core state model' while others have used the LDA (GGA) + $U$
formalism, where an onsite Coulomb repulsion is incorporated to account for
the repulsive energy arising from the double occupancy of an $f$ orbital on a
Ce site. \cite{Andersson2007}

The problem of the $4f$ states in the ground state configurations of CeO$_{2}$
and Ce$_{2}$O$_{3}$ resembles the difficulties also encountered in studying
these same states in Ce metal. Ce metal undergoes an isostructural $\alpha
$-$\gamma$ phase transition with a $15\%$ change in volume.
\cite{Eriksson1991} The average $f$ electron populations observed
experimentally are $0.81$ and $0.97$ for $\alpha$- and $\gamma$-Ce
respectively. \cite{Rueff2006} Thus, the pure Ce metal shows significant mixed
valence in the $\gamma$ phase.

\section{The Bond Valence Model}

\label{sec:BVM}

The bond valence model (BVM), which is a generalization of Pauling's original
concept of the electrostatic valence principle, has been reviewed extensively
in the literature \cite{Brown1992,Brown2002,Burdett1995a}. A quantum chemical
justification of the model has been discussed by Mohri. \cite{Mohri2003} The
model defines the relationship between bond valences, $s$, and the
corresponding atomic valences, $V$, through the equations:%
\begin{equation}
V_{i}=%
{\textstyle\sum\limits_{j}}
s_{ij} \label{bond 01}%
\end{equation}
where $i$ and $j$ refer to different atoms. This relationship is called the
valence-sum rule.

There is a well-defined empirical relationship between bond valences as
defined in (\ref{bond 01}) and bond lengths in a given coordination polyhedron
and it is this functional relationship which makes the BVM quantitatively
useful. The relationship is monotonic, and over the small range in which most
bonds are found, it can be approximated by: \cite{Brown2002}%
\begin{equation}
s_{ij}=\exp\left(  \frac{R_{0}-R_{ij}}{B}\right)  \label{bond 02}%
\end{equation}
Here $R_{0}$ and $B$ are fitted parameters, $R_{0}$ being the bond length of a
bond of unit valence. For a specific bond, these parameters are determined by
fitting the measured lengths of bonds in a wide range of compounds by
enforcing the valence-sum rule (\ref{bond 01}). It has been shown that for
most bonds, $B$ can actually be set to be $0.37%
\operatorname{\text{\AA}}%
$ which reduces the bond valence model (\ref{bond 02}) to a one-parameter
model. \cite{Brese1991} In that case, for each structure where a central atom
is bonded only to atoms of a particular species, the $R_{0}$ parameter is then
obtained by combining the valence-sum rule (\ref{bond 01}) and (\ref{bond 02}%
), i.e:%
\begin{align}
V_{i}  &  =%
{\textstyle\sum\limits_{j}}
s_{ij}=%
{\textstyle\sum\limits_{j}}
\exp\left(  \frac{R_{0i}-R_{ij}}{B}\right) \nonumber\\
&  =\exp\left(  \frac{R_{0i}}{B}\right)
{\textstyle\sum\limits_{j}}
\exp\left(  \frac{-R_{ij}}{B}\right)  \label{bond 04a}%
\end{align}
Which can be rewritten in the form:%

\begin{equation}
R_{0i}=B\ln\left[  \frac{V_{i}}{%
{\textstyle\sum\limits_{j}}
\exp\left(  \frac{-R_{ij}}{B}\right)  }\right]  \label{bond 05}%
\end{equation}
The BVM is applicable to \emph{bipartite} crystals which, in our case, means
that only Ce-O bonds can exist in any given crystal. In all cases, the
bipartite requirement is satisfied for Ce polyhedra considered below. However
this is not true of all O polyhedra as in a few cases, e.g., in Ce$_{6}%
$O$_{11}$, O atoms are included in the coordination polyhedron of an O atom.
Nevertheless, this does not concern us as all our calculations are performed
on cation-centred polyhedra. A second issue arises from how to precisely
define the coordination number for some of the Ce sites. We adopt Brown's
definition of the coordination number as the number of atoms (O, in this case)
to which a central atom (Ce, in this case) is bonded. \cite{Brown2002a} The
operational meaning of this definition was given by Altermatt and
Brown:\cite{Altermatt1985} a bond exists between a cation and an anion if its
experimental bond valence is larger than $0.04$ $\times$ the cation valence.

Various authors have determined values for the parameters $R_{0}$ and $B$ in
Eq. (\ref{bond 02}) for Ce-O bonds from an analysis of measured bond
lengths in both organic \cite{Roulhac2003,Trzesowska2004} and inorganic
compounds. \cite{Brese1991,Zocchi2007} It has been reported that Ce-O bonds in
inorganic compounds are longer than in metal-organic coordination compounds
and so the bond parameters are larger \cite{Trzesowska2004}. Although Zocchi
has derived detailed parameters for the Ce-O bonds in iorganic compounds
explicitly giving the dependence on coordination number \cite{Zocchi2007}, his
so-called method of intercepts \cite{Zocchi2002} does not provide a clear
physical basis for the calculation. In view of these considerations, we
selected the parameterization of the BVM by Brese and O'Keeffe.
\cite{Brese1991}

In this parameterization, $B=0.37%
\operatorname{\text{\AA}}%
$ with $R_{0}=2.151%
\operatorname{\text{\AA}}%
$ and $R_{0}=2.028%
\operatorname{\text{\AA}}%
$ for Ce$^{3+}$ and Ce$^{4+}$ ions respectively. This means that, by Brown's
criterion for the coordination sphere, only O atoms within the critical
distances $R_{c}=2.905%
\operatorname{\text{\AA}}%
$ and $R_{c}=2.746%
\operatorname{\text{\AA}}%
$ are included within the coordination polyhedron for Ce$^{3+}$ and Ce$^{4+} $
sites respectively.

\begin{figure}
[ptb]
\begin{center}
\includegraphics[
natheight=5.333300in,
natwidth=6.106400in,
height=2.6948in,
width=3.0805in
]%
{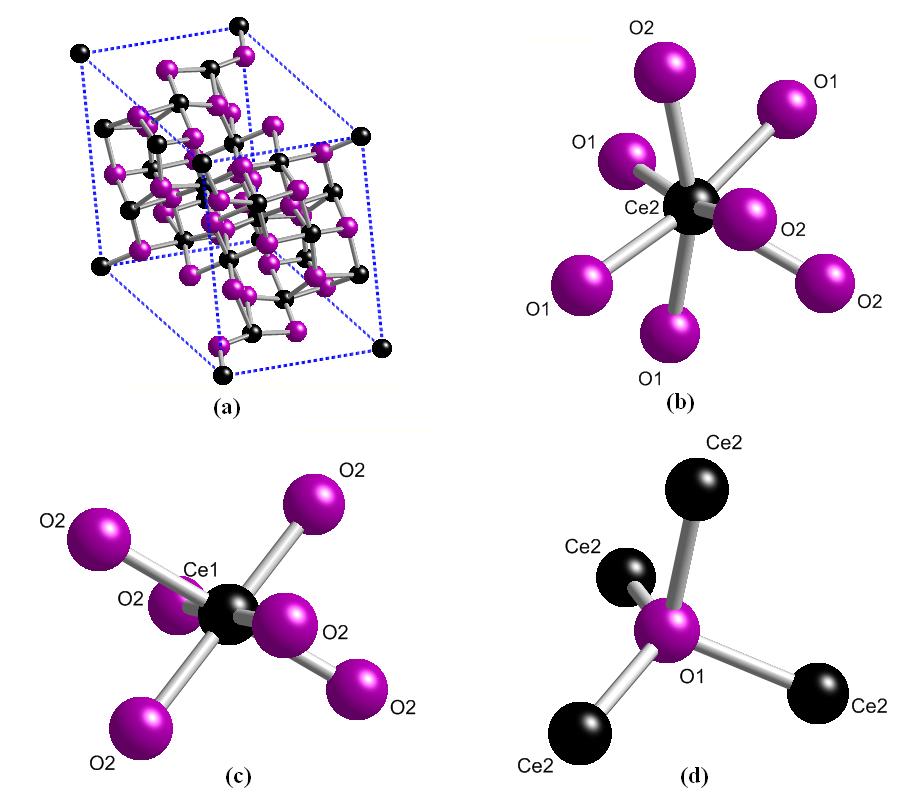}%
\caption{Identification of the coordination polyhedra in Ce$_{7}$O$_{12}$ for
bond valence calculations. Atom labels refer to the distinct sites in the
crystal. (a) The unit cell of Ce$_{7}$O$_{12}$ in the hexagonal crystal
lattice. The space group is $R3$ and the unit cell consists of three formula
units. (b) The polyhedron of the Ce(2) site which is seven-coordinated and of
triclinic symmetry and (c) that of the the six-coordinated Ce(1) site of
S$_{6}$ symmetry. There are two distinct polyhedra for the O atoms and one of
these, the O(1) site is shown in (d). In this polyhedron, the O atom has a
coordination of four in triclinic symmetry. The colours of the atoms are: Ce -
black and O - violet. Images generated in CrystalMaker \cite{CrystalMaker}.}%
\label{polyhedron}%
\end{center}
\end{figure}

To perform bond valence calculations, the only input required is the bond
length data of the Ce-O bonds in the respective coordination polyhedra of the
various oxides. For this, we obtained crystallographic data from the sources
listed in Table \ref{valency}. For each oxide, all the distinct coordination
spheres for both Ce and O atoms were identified. Fig. \ref{polyhedron}
illustrates the procedure using the example of Ce$_{7}$O$_{12}$ where we have
shown only one of the two O atom polyhedra in the crystal of this oxide. The
Ce-O bond distances in each polyhedron were then obtained from the
crystallographic data using CrystalMaker. \cite{CrystalMaker} We have listed
the Ce-O bond length data in Appendix A. We note here two oxides, $C$-Ce$_{2}%
$O$_{3}$ and Ce$_{6}$O$_{11}$, for which we were not able to obtain full
crystallographic data. $C$-Ce$_{2}$O$_{3}$ has not been observed
experimentally and the lattice constant used here, $11.22%
\operatorname{\text{\AA}}%
$, was derived by Eyring \cite{Eyring1979} and also by Tsunekawa \textit{et
al.}. \cite{Tsunekawa1999} We could not find positional parameters for the
crystal of Ce$_{6}$O$_{11}$ and we used the positional parameters for Pr$_{6}%
$O$_{11}$ with which it is isostructural. \cite{Sorensen1976}%

\bigskip

\section{Results and Discussion}

\label{sec:results}

We have noticed from the results of Roulhac \cite{Roulhac2003} that the
parameter $R_{0}$ varies approximately linearly with the oxidation state of
the Ce atom between $R_{0}=2.121(13)%
\operatorname{\text{\AA}}%
$ (Ce$^{3+}$) and $R_{0}=2.068(12)%
\operatorname{\text{\AA}}%
$ (Ce$^{4+}$). We exploit this relationship between $R_{0}$ and the Ce
oxidation state to calculate the $f$ occupancies of the Ce sites
self-consistently in the following way. As already mentioned, the valence at
each Ce site fluctuates between the $f^{0}$ (Ce$^{4+}$) and $f^{1}$ (Ce$^{3+}%
$) configurations. Let $x$ be the probability that a Ce ion is in the $f^{1}$
configuration $\left(  0\leq x\leq1\right)  $, then, the corresponding
probability of the $f^{0}$ configuration at the same site is given by $1-x$.
Thus, for such a mixed-valence Ce site, the correponding value of the
parameter $R_{0}$ is then given by linear interpolation between these values:%
\begin{equation}
R_{0}=R_{3}^{0}x+R_{4}^{0}\left(  1-x\right)  \label{bond 06}%
\end{equation}
As already mentioned, we have used $R_{3}^{0}=2.151%
\operatorname{\text{\AA}}%
$ and $R_{4}^{0}=2.028%
\operatorname{\text{\AA}}%
$, i.e., $R_{0}$\ parameters for Ce$^{3+}$ and Ce$^{4+}$ states respectively.
\cite{Brese1991} From (\ref{bond 06}) and (\ref{bond 02}), we have:%
\begin{equation}
s_{ij}\left(  x\right)  =\exp\left(  \frac{\left[  R_{3}^{0}x+R_{4}^{0}\left(
1-x\right)  \right]  -R_{ij}}{B}\right)  \label{bond 07}%
\end{equation}
It then follows that, for a Ce site $i$, $V_{i}\left(  x\right)  =4-x=%
{\textstyle\sum\limits_{j}}
s_{ij}\left(  x\right)  $, which then with $N_{f}=x$, leads to the required
self-consistent equations:%

\begin{equation}
x=4-%
{\textstyle\sum\limits_{j}}
\exp\left(  \frac{\left[  R_{3}^{0}x+R_{4}^{0}\left(  1-x\right)  \right]
-R_{ij}}{B}\right)  \label{bond 09}%
\end{equation}
We solved (\ref{bond 09}) for $x=N_{f}$ to get the $f$ occupancies of the
various Ce sites in the seven crystals studied. The results of applying
Eq. (\ref{bond 09}) to the seven oxides of cerium selected for this study
are given in Table \ref{valency}.%

\begin{table*}[htb] \centering
\caption{Site valencies of Ce at different sites in cerium oxides (CeO$_{2-y}$) calculated from our self-consistent bond valence
method.}%
\begin{tabular}
[c]%
{p{0.65in}p{0.65in}p{0.65in}p{0.65in}p{0.65in}p{0.65in}p{0.65in}p{0.65in}p{0.65in}p{0.65in}}%
\hline\hline
Oxide & $y$ & Ref. & Ce site & Coord. No. & Site Symmetry & No. of Sites in
Unit Cell & Site Bond Valence Sum (v.u) & $N_{f}$ & $\%$ Variation\\\hline
$A$-Ce$_{2}$O$_{3}$ & $0.50$ & \onlinecite{Barnighausen1985,Wyckoff1964} &
Ce & $7$ & $C_{3v}$ & $2$ & $2.97$ & $1.03$ & $14$\\
$C$-Ce$_{2}$O$_{3}$ & $0.50$ &
\onlinecite{Wyckoff1964,Villars1991,Kummerle1999} & Ce(1) & $6$ & $S_{6}$ &
$8$ & $2.98$ & $1.02$ & $0.0$\\
&  &  & Ce(2) & $6$ & $C_{2}$ & $24$ & $2.98$ & $1.02$ & $4$\\
Ce$_{3}$O$_{5}$ & $0.33$ &
\onlinecite{Kummerle1999,Zinkevich2006,Wyckoff1964} & Ce(1) & $8$ & $S_{6}$ &
$8$ & $3.33$ & $0.67$ & $8.5$\\
&  &  & Ce(2) & $8$ & $C_{2}$ & $24$ & $3.37$ & $0.63$ & $8.7$\\
Ce$_{7}$O$_{12}$ & $0.29$ & \onlinecite{Bartram1966,Ray1975} & Ce(1) & $6$ &
$S_{6}$ & $3$ & $3.67$ & $0.33$ & $0.0$\\
&  &  & Ce(2) & $7$ & $\bar{1}$ & $18$ & $3.21$ & $0.79$ & $7.0$\\
Ce$_{11}$O$_{20}$ & $0.18$ & \onlinecite{Kummerle1999,Zhang1993} & Ce(1) & $8$
& $\bar{1}$ & $2$ & $3.08$ & $0.92$ & $13$\\
&  &  & Ce(2) & $8$ & $\bar{1}$ & $2$ & $3.06$ & $0.94$ & $7.6$\\
&  &  & Ce(3) & $7$ & $\bar{1}$ & $2$ & $3.46$ & $0.54$ & $7.8$\\
&  &  & Ce(4) & $7$ & $\bar{1}$ & $2$ & $3.58$ & $0.42$ & $12.5$\\
&  &  & Ce(5) & $7$ & $\bar{1}$ & $2$ & $3.68$ & $0.32$ & $8.2$\\
&  &  & Ce(6) & $7$ & $\bar{1}$ & $2$ & $3.76$ & $0.24$ & $9.4$\\
Ce$_{6}$O$_{11}$ & $0.17$ & \onlinecite{Sorensen1976,Villars1991,Zhang1996} &
Ce(1) & $6$ & $1$ & $4$ & $3.22$ & $0.78$ & $30.9$\\
&  &  & Ce(2) & $7$ & $\bar{1}$ & $4$ & $3.75$ & $0.25$ & $26.7$\\
&  &  & Ce(3) & $7$ & $\bar{1}$ & $4$ & $3.10$ & $0.90$ & $15.1$\\
&  &  & Ce(4) & $5$ & $\bar{1}$ & $4$ & $3.62$ & $0.38$ & $37.4$\\
&  &  & Ce(5) & $6$ & $\bar{1}$ & $4$ & $2.93$ & $1.07$ & $24.0$\\
&  &  & Ce(6) & $6$ & $\bar{1}$ & $4$ & $3.22$ & $0.78$ & $30.9$\\
CeO$_{2}$ & $0.00$ & \onlinecite{Villars1991,Taylor1984,Sorensen1976} & Ce &
$8$ & $O_{h}$ & $4$ & $3.73$ & $0.27$ & $0.0$\\\hline\hline
\end{tabular}
\label{valency}%
\end{table*}%

For CeO$_2$, all O atoms are symmetry equivalent (as are all Ce atoms).  For this simple case, we can write down an equation describing the polyhedron centred on given O, equivalent to (\ref{bond 09}), and use it to determine the valence of the O atom self-consistently.  The valence on the oxygen atom of CeO$_2$ is given by:%
\begin{align}
V_{i}\left(  x\right)   &  =-1.5x-2(1-x)=%
{\textstyle\sum\limits_{i=1}^{4}}
s_{ji}\left(  x\right)  =-%
{\textstyle\sum\limits_{i=1}^{4}}
s_{ij}\left(  x\right) \nonumber\\
&  =-%
{\textstyle\sum\limits_{i=1}^{4}}
\exp\left(  \frac{\left[  R_{3}^{0}x+R_{4}^{0}\left(  1-x\right)  \right]
-R_{ij}}{B}\right) \label{bond 11aa}\\
&  \Rightarrow x=4-2%
{\textstyle\sum\limits_{i=1}^{4}}
\exp\left(  \frac{\left[  R_{3}^{0}x+R_{4}^{0}\left(  1-x\right)  \right]
-R_{ij}}{B}\right)  \label{bond 11b}%
\end{align}
Evaluation of (\ref{bond 11b}) gives $x=0.268$ from where it follows that the
O valence is $-1.87$ which is consistent with the Ce valence already
calculated above.  

In the general case, in which there are inequivalent O atoms, we cannot determine the valence of a given O site in this way.  That is, while the valence of each Ce atom can be determined by considering the polyhedron consisting of this central Ce and its nearest neighbors, the same is not generally true of an O atom.  This is because we have assumed that the parameter $R_0$, a property of a Ce-O bond, depends on the valence of the Ce atom but not on the valence of the O atom.  This assumption is consistent with the observation that the structures of different Ce-oxides with the same Ce-valence can be adequately described by (\ref{bond 02}) using a single value for $R_0$ (Ref.\onlinecite{Brown1981}).           

Locock and Burns have defined a measure of the variation in bond lengths of
the bonds included in a coordination polyhedron as follows: \cite{Locock2004}%
\begin{equation}
variation(\%)=\frac{\left\vert bond_{\max}-bond_{\min}\right\vert }%
{bond_{avg}}\cdot100 \label{vary}%
\end{equation}
where the subscripts $\max$, $\min$, and $avg$ refer to the maximum, minimum
and average bond lengths in the polyhedron. We have calculated the $\%$
variation for all the polyhedra in this study and the results are given in
Table \ref{valency}.

We now give a few remarks about the accuracy of the bond valence method. Brown
has estimated that bond valence sums have an accuracy of $0.05$ v.u.
\cite{Brown1981} The main error in the method comes from the fitted parameters
of the model ($B$, $R_{0}$). It has been noted that bond valence parameters
which overestimate valences of strong bonds while underestimating those weak
bonds will give bond valence sums which are too high in low coordination
polyhedra and too low in the case of high coordination. \cite{Krivovichev2001}
We have estimated that with the following uncertainties in the model
parameters and bond length data: $R_{0}$ ($\pm0.01%
\operatorname{\text{\AA}}%
$), $R_{ij}$ ($\pm0.01%
\operatorname{\text{\AA}}%
$) \cite{Brese1991} and $B$ ($0.037%
\operatorname{\text{\AA}}%
$) \cite{Brown2002} the uncertainties in the bond valence sums for CeO$_{2}$
and $A-$Ce$_{2}$O$_{3}$ are $\pm0.13$ vu and $\pm0.12$ v.u respectively which
is in agreement with Ref. \onlinecite{Liebau2005}.

\begin{figure}
[ptb]
\begin{center}
\includegraphics[
natheight=6.653000in,
natwidth=9.346900in,
height=2.3566in,
width=3.2993in
]%
{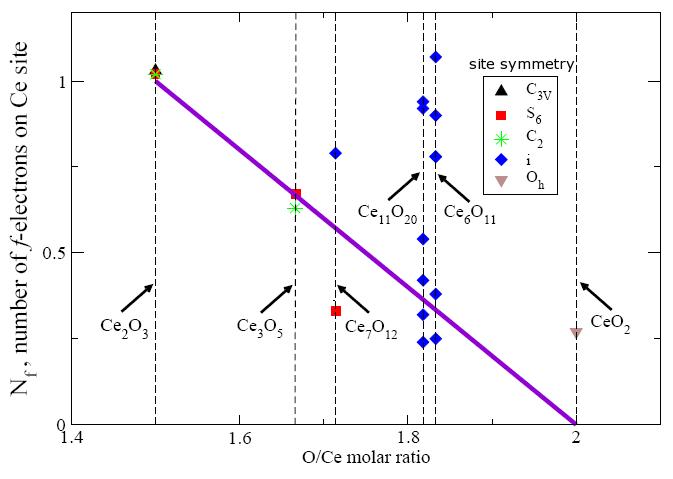}%
\caption{Ce $f$-level site occupancies in different crystallographic phases of
the oxides calculated self-consistently in the bond valence model. The results
are given according to the exact point group symmetries of the respective Ce
sites in each crystallographic phase. As can be seen from the legend, the
point symmetries of the Ce sites vary from as low as triclinic (i) to as high
as octahedral (O$_{h}$), the point group of the cube. The straight line
represents $N_{f}$ values obtained from a simple Ce valence calculation based
on the electroneutrality of the formula unit and the assumption of an
oxidation state of $-2$ for the O ions in all the oxides.}%
\label{occupancies}%
\end{center}
\end{figure}

In Fig. \ref{occupancies}, we have plotted the Ce site $f$ occupancies,
$N_{f}$, calculated from Eq. (\ref{bond 09}) . The results are plotted
for increasing average degree of oxidation of the Ce ion. For each crystal,
the results are reported according to the site symmetries of the Ce sites
which range from the lowest triclinic sites ($i$) to the most symmetrical
octahedral sites ($O_{h}$). The dashed line labelled Ce$_{2}$O$_{3}$ refers to
both the $A$- and $C$-phases of this oxide. For comparison, we have included
the solid straight line which represents $N_{f}$ values calculated from the
formula units by requiring charge balance of the formula unit, a valence of
$-2$ for all O ions and an even distribution of the valence among all the Ce
ions in a formula unit. We will call this method of calculating $N_{f}$ the
homogeneous mixed-valence approximation (HMA) as it assigns the same valence
to each cationic site regardless of the specific site properties. We contrast
this method with another simple approach for predicting cationic valences
which we will call the inhomogeneous mixed-valence approximation (IHMA). The
IHMA is based on the requirement that all valences must be integral but their
exact assignment in the crystal lattice does not necessarily have to satisfy
constraints which may arise from specific differences in local site symmetry.
Thus, this approximation considers the crystal lattice to be a static charge
ordered array.%

The results plotted in Fig. \ref{occupancies} show that mixed-valence in Ce
oxides does not fit nicely into either of the traditional classes of
mixed-valence, i.e., homogeneous\emph{ }or inhomogeneous mixed-valence as
originally defined by Varma. \cite{Varma1976} For the Ce oxides, only
CeO$_{2}$ and Ce$_{2}$O$_{3}$ are strictly homogeneous mixed-valent oxides
since all the Ce sites are symmetrically equivalent and have the same
oxidation state. The rest of the oxides do not have symmetrically equivalent
Ce sites and the oxidation states of the individual sites are generally
different within the error limits of the method. The mixed-valence in the
crystal is not a result of the averaging of the oxidation states at the
different sites but a property of the individual sites, which means that the
mixed-valence is not of the simple inhomogeneous type.

It is interesting and important to compare our results to those obtained by
other methods. In Table \ref{refs} we have reported the results of the Ce site
$4f$ occupancy, $N_{f}$, from the literature. We have not been able to obtain
results for all other oxides except for CeO$_{2}$ and Ce$_{2}$O$_{3}$. The
data in Table \ref{refs} shows that while there is good agreement on the $f$
occupancy of the Ce site in $A$-Ce$_{2}$O$_{3}$, there is considerable
variation in the reported results for CeO$_{2}$. The $N_{f}$ values for
CeO$_{2}$ vary from $0.00$ for LDA and GGA calculations to $0.60$ for Ce $3d$
core XPS. We now discuss our results in detail for the individual oxides. The
format of our discussion is as follows: for each oxide, we compare the
predicted valences from the BVM to those obtainded from HMA and IHMA. We then
indicate whether the particular oxide exhibits mixed-valence or not. We sum up
this section by relating our results to the conventional model of vacancy
formation and ordering in bulk Ce oxides and the possible role of electron
correlation in these oxides.%

\begin{table*}[htb] \centering
\caption{$N_{f}$ values for the Ce ion in cerium oxides determined by different methods}\begin{minipage}{\textwidth}
\setcounter{mpfootnote}{\value{footnote}}
\renewcommand{\thempfootnote}{\arabic{mpfootnote}}
\begin{tabular}
[c]{p{1in}p{1in}p{1.8in}p{1.8in}p{1in}}\hline\hline
Compound & $N_{f}$ & Method & Remarks & Ref.\\\hline
CeO$_{2}$ & $0.3\pm0.1$ & Bond valence & From crystal structure data & This
work\\
& $0.60$ & Ce $3d$ core XPS & Bulk measurement & \onlinecite{Fujimori1983}\\
& $0.45$ & XANES & Bulk measurement & \onlinecite{Wu2001}\\
& $0.50$ & Ce $3d$ core XPS & Bulk measurement & \onlinecite{Kotani1988}\\
& $0.05$ & Optical reflectance & Bulk, measurement &
\onlinecite{Marabelli1987}\\
& $0.2-0.4$ & LDA $+U$ & Bulk, calculation & \onlinecite{Castleton2007}\\
& $0.50$ & LAPW $\chi_{\alpha}$ & Bulk, calculation &
\onlinecite{Koelling1983}\\
& $0.40$ & Ce $3d$ XAS & Bulk, calculation & \onlinecite{Kotani1989}\\
& $0.20$ & HSE & Bulk, calculation & \onlinecite{Hay2006}\\
& $0.00$ & LDA and GGA & Bulk, calculation & \onlinecite{Skorodumova2001}\\
Ce$_{2}$O$_{3}$ & $1.0\pm0.1$ & Bond valence & From crystal structure data &
This work\\
& $1.0$ & Ce $3d$ core XPS & Bulk measurement & \onlinecite{Kotani1988}\\
& $1.0$ & LDA and GGA\footnote{Both the LDA and GGA calculations were performed by artificially localizing the $4f$ states, the so called core state model. Hence, ${N_{f}=1}$ was actually assumed in the calculation.}& Bulk, calculation. & \onlinecite{Skorodumova2001}\\
& $1.0$ & HSE & Bulk, calculation & \onlinecite{Hay2006}\\
Ce$_{2}$Zr$_{2}$O$_{7.5}$ & $0.5$ & EELS (Ce-M$_{4,5}$) & Bulk measurement,
ceria-zirconia solid solution & \onlinecite{Arai2004}\\\hline\hline
\end{tabular}
\setcounter{footnote}{\value{mpfootnote}}
\end{minipage}
\label{refs}%
\end{table*}%

\subsection{$A$-Ce$_{2}$O$_{3}$}

The bond valence method gives $N_{f}=1.0$ for this oxide. As can be seen from
Table \ref{refs}, the bond valence method gives $N_{f}$ values which accord
well with results from the other methods for this oxide. In addition, both the
HMA (Fig. \ref{occupancies}) and IHMA also predict $N_{f}=1.0$ and so all the
methods are in agreement. Compared to CeO$_{2}$, Table \ref{refs} shows that
the variation in bond lengths in the Ce polyhedron of $A$-Ce$_{2}$O$_{3}$ is
relatively high at about $14\%$. However, the result of $N_{f}$ obtained in
this calculation which is in good agreement with other methods may suggest
that bond lengths distortions of this magnitude may have no significant role
in the bond valence model. We conclude that mixed-valence in $A$-Ce$_{2}%
$O$_{3}$ is negligible.

\subsection{$C$-Ce$_{2}$O$_{3}$}

The results are similar to those for $A$-Ce$_{2}$O$_{3}$ just described above
in both the $S_{6}$ and $C_{2}$ site symmetries of $C$-Ce$_{2}$O$_{3}$.
However, structurally, the two sesquioxides are very different. The
$A$-sesquioxide has a hexagonal Bravais lattice (space group $P\bar{3}m1$)
with a lattice constants $a=3.891%
\operatorname{\text{\AA}}%
$ and $b=6.059%
\operatorname{\text{\AA}}%
$ and only one Ce site of $C_{3v}$ symmetry. In contrast, the $C$-sesquioxide
has a cubic fluorite structure with two O vacancies along the body and face
diagonals (space group $Ia\bar{3}$) giving the $S_{6}$ and $C_{2}$ site
symmetries for the Ce ions respectively. That the Ce sites in $A$-Ce$_{2}%
$O$_{3}$ and $C$-Ce$_{2}$O$_{3}$ still have the same $N_{f}$ value despite the
differences in point symmetry appears to suggest that site symmetry may have
no significant influence on the valence of a given site in Ce$_{2}$O$_{3}$.
Compared to the other oxides included in this study, the only other Ce site
with a comparable $N_{f}$ value is the triclinic Ce(5) site in Ce$_{6}$%
O$_{11}$.

\subsection{Ce$_{3}$O$_{5}$}

The results in Table \ref{valency} indicate that both Ce sites in this crystal
have comparable valences with $N_{f}$ values of $0.67$ and $0.63$ for the
$S_{6}$ and $C_{2}$ sites respectively. Again, this result highlights the
earlier observation that site symmetry may have no significant role in site
valences as the two $N_{f}$ values of these two sites are comparable. The HMA
predicts that $N_{f}=0.67$ for this crystal which compares favourably with the
BVM while the IHMA requires the ionic formulation\linebreak\ $\left[  \left(
\text{Ce}^{3+}\right)  _{2}\text{Ce}^{4+}\right]  \left(  \text{O}%
^{2-}\right)  _{5}$ for Ce$_{3}$O$_{5}$. Thus, the IHMA indicates that in a
static charge-ordered array of integral valences, the Ce$_{3}$O$_{5}$ lattice
has twice as many Ce$^{3+}$ ions as there are Ce$^{4+}$ ions. Although the
Ce(1) and Ce(2) sites are of different point group symmetries, they are of
comparable polyhedron sizes with average Ce-O distances of $2.444%
\operatorname{\text{\AA}}%
$ and $2.435%
\operatorname{\text{\AA}}%
$ respectively. When one considers how to distribute the Ce$^{3+}$ and
Ce$^{4+}$ ions between these sites, there is a conceptual difficulty. Firstly,
in typical ionic crystals of Ce, the average radii of the ions in an
$8$-coordinate environment are $1.28%
\operatorname{\text{\AA}}%
$ and $1.11%
\operatorname{\text{\AA}}%
$ for Ce$^{3+}$ and Ce$^{4+}$ respectively. \cite{Shannon1976} This $13\%$
difference in the crystal radii of the ions is significantly different from
the $0.4\%$ difference in the average sizes of the coordination polyhedra.
Secondly, the ratio of the Ce(1) and Ce(2) sites does not match that of the
ions as given in the ionic formulation above. Thus, if one were to argue that
the smaller Ce$^{4+}$ ion will show a slight preference for the smaller Ce(2)
site, then all the Ce$^{4+}$ ions would occupy the Ce(2) sites. However, since
there are almost twice as many Ce(2) sites as there are Ce$^{4+}$ ions, the
remainder of these sites will be occupied by the Ce$^{3+}$ ions and the
latter, will, in addition, occupy all the Ce(1) sites. This, of course,
results in a situation where Ce$^{3+}$ and Ce$^{4+}$ occupy the same type of
site in the crystal. This logical inconsistency does not arise when the BVM is
applied and we conclude that Ce$_{3}$O$_{5}$ is a homogeneous mixed-valence compound.

\subsection{Ce$_{7}$O$_{12}$}

Both the Ce(1) and Ce(2) sites are predicted in the BVM to be mixed-valent
with the more symmetrical Ce(1) site closer to Ce$^{4+}$ ($3.67$ v.u) and the
triclinic Ce(2) site closer to Ce$^{3+}$ ($3.21$ v.u) whereas the HMA gives a
valence of $3.4$ v.u. The ionic formulation for the IHMA is $\left[  \left(
\text{Ce}^{3+}\right)  _{4}\left(  \text{Ce}^{4+}\right)  _{3}\right]  \left(
\text{O}^{-2}\right)  _{12}$ giving the unit cell, $\left[  \left(
\text{Ce}^{3+}\right)  _{12}\text{Ce}_{9}^{4+}\right]  \left(  \text{O}%
^{-2}\right)  _{36}$ for Ce$_{7}$O$_{12}$. The Ce(1) site is smaller than the
Ce(2) site and it has been suggested that since the lower coordination at the
Ce(1) site would require a smaller cation, then Ce$^{4+}$ ions are expected to
occupy all three of these sites. The remaining six Ce$^{4+}$ and the twelve
Ce$^{3+}$ then occupy the Ce(2) which leads to a similar problem as already
noted for Ce$_{3}$O$_{5}$ above. When the BVM and IHMA results are compared,
we notice that the BVM predicts that there are two distinct valences for the
Ce ions, one closer to Ce$^{3+}$ and the other closer to Ce$^{4+}$but without
the inconsistency of assigning different valences to the same type of Ce site.
These results indicate mixed-valence in Ce$_{7}$O$_{12}$.

\subsection{Ce$_{11}$O$_{20}$}

This oxide has the lowest symmetry of all the cerium oxides included in this
study. Based on the the BVM results in Table \ref{valency}, the following
approximate assignments of site oxidation states can be made: Ce(1) and Ce(2)
- Ce$^{3+}$, Ce(3) and Ce(4) - strongly mixed-valent, Ce(5) and Ce(6) - closer
to Ce$^{4+}$. Fig. \ref{occupancies} shows that the HMA predicts a valence of
$3.6$ v.u. which is also mixed-valent. The ionic formulation $\left[  \left(
\text{Ce}^{3+}\right)  _{4}\left(  \text{Ce}^{4+}\right)  _{7}\right]  \left(
\text{O}^{-2}\right)  _{20}$ results from applying the IHMA to this crystal.
Both the BVM and the IHMA predict the presence of Ce$^{3+}$ sites in this
crystal and a consistent distribution of the sites is obtained between the
methods since it is expected that, in the IHMA ionic formulation, the larger
Ce$^{3+}$ ion will occupy the larger eight-coordinated polyhedra Ce(1) and
Ce(2) which is the same result obtained from the BVM analysis. However, the
two methods disagree on the assignment of valencies to the remainder of the
sites with the IHMA assigning them all to Ce$^{4+}$ while the BVM clearly
indicates strong mixed-valence for the Ce(3) and Ce(4) sites and less but
still significant mixed valence for the Ce(5) and Ce(6) sites. The HMA does
not appear to be a viable proposition for this crystal lattice because of its
very low symmetry. Again, we find that Ce$_{11}$O$_{20}$ is a mixed-valence
compound but one that does not fit into either of the traditional classes.

\subsection{Ce$_{6}$O$_{11}$}

The results in Table \ref{valency} obtained from the BVM show that all Ce
sites in Ce$_{6}$O$_{11}$ are mixed-valent with deviations from the nearest
integral valences increasing from $0.1$ v.u. to $0.38$ v.u. in the order
Ce(5), Ce(3), Ce(1), Ce(6), Ce(2) and Ce(4). As shown in Fig \ref{occupancies}%
, the HMA predicts mixed-valence for this compound giving a valence of $3.7$
v.u for the Ce ions. Considered in the IHMA, the ionic composition of Ce$_{6}%
$O$_{11}$ should be $\left[  \left(  \text{Ce}^{3+}\right)  _{2}\left(
\text{Ce}^{4+}\right)  _{4}\right]  \left(  \text{O}^{2-}\right)  _{11}$ which
implies that there are twice as many Ce$^{4+}$ ion sites as there are
Ce$^{3+}$ sites in a unit cell of Ce$_{6}$O$_{11}$. We notice that the
predictions of the three methods are very different for this crystal. Table
\ref{valency} shows that this crystal has the most distorted polyhedra with
$\%$ variation in the bond lengths ranging $15\%-37\%$ and it is expected that
the BVM would reflect this aspect of the local site geometries. With these
complex local geometries, it is not expected that the HMA would give a good
approximation to the site valences. Comparing the IHMA and the BVM, we notice
that even if one tentatively considered both Ce(2) and Ce(4) to be Ce$^{4+}$
sites and the rest Ce$^{3+}$, then one gets twice as many Ce$^{3+}$ sites as
there are Ce$^{4+}$ sites. This charge-ordering is the reverse of that
predicted by the IHMA ionic formulation which clearly highlights the
disagreement between the methods.

\subsection{CeO$_{2}$}

Our value of $N_{f}=0.27$ for CeO$_{2}$ is comparable to the result obtained
by Castleton \textit{et al.} \cite{Castleton2007} from an $LDA+U$ calculation
as shown in Table \ref{refs}. We can see from Table \ref{refs} that results
from $3d$ core-level spectroscopy tend to give relatively high $N_{f}$ vaues
($N_{f}\geq0.40$). In CeO$_{2}$, the Ce site is in a symmetric polyhedron with
all the Ce-O bonds equal in length and therefore the bond valence model is
expected to perform well for this oxide. As can be seen from Fig.
\ref{occupancies}, the HMA predicts that the Ce ion is in a pure $f^{0}$
configuration in CeO$_{2}$. The same result is obtained from the IHMA and,
thus, both methods contradict the BVM. From our result, we conclude that
CeO$_{2}$ is a mixed-valent compound.

It would be informative to compare the bond valence sum results obtained here
to those for their Pr oxide counterparts. However, we have not found any
detailed study of the bond valence sums of Pr oxides for a meaningful
comparison to be made here. We only found a bond valence calculation performed
for the Pr-O bond in the high-T$_{c}$ superconductor PrYBa$_{2}$Cu$_{3}$%
O$_{7}$ where the Pr was mixed-valent with a valence of $3.4$ v.u.
\cite{Guillaume1993}

\section{Conclusion}

\label{sec:conclusion}

Since we did not obtain integral $N_{f}$ values for most of the Ce oxide
sites, our results suggest that mixed valence is an essential feature of Ce in
its pure oxide phases. We have also shown that for a given crystallographic
phase, several different oxidation states may exist for Ce sites which belong
to the same point group. Thus, it appears that valence fluctuations depend on
the exact coordination geometry of a Ce site which suggests that valence
fluctuation is a `local' property of the Ce site. The point group symmetry of
a Ce site does not appear to play a significant role in determining the
valence state of that site. We have also noted that the bond valence method
eliminates some of the conceptual difficulties which arise when one attempts
to distribute Ce ions of different integral valences in sites of the same type
in a given crystal.

\begin{acknowledgments}
Our interest in the study of cerium oxides was motivated by our discussions
with Prof. C. Stampfl at the University of Sydney. We thank B. Powell, A.
Jacko, E. Scriven, J. Merino, M. Yethiraz, A. Stilgoe, B. Mostert and H. F.
Schaefer for helpful discussions. One of us (E. S) is grateful to the
Australian Commonwealth Government Department of Science Education and
Training for the award of the International Postgraduate Research Scholarship
(IPRS) and to the University of Queensland for the University of Queensland
International Postgraduate Research Scholarship (UQIPRS). This work was also
supported by the Australian Research Council.
\end{acknowledgments}

\bibliographystyle{apsrev}
\bibliography{prb}

\begin{thebibliography}{89}
\expandafter\ifx\csname natexlab\endcsname\relax\def\natexlab#1{#1}\fi
\expandafter\ifx\csname bibnamefont\endcsname\relax
  \def\bibnamefont#1{#1}\fi
\expandafter\ifx\csname bibfnamefont\endcsname\relax
  \def\bibfnamefont#1{#1}\fi
\expandafter\ifx\csname citenamefont\endcsname\relax
  \def\citenamefont#1{#1}\fi
\expandafter\ifx\csname url\endcsname\relax
  \def\url#1{\texttt{#1}}\fi
\expandafter\ifx\csname urlprefix\endcsname\relax\def\urlprefix{URL }\fi
\providecommand{\bibinfo}[2]{#2}
\providecommand{\eprint}[2][]{\url{#2}}

\bibitem[{\citenamefont{{For a recent survey, see Mixed Valency: Papers of a
  Discussion Meeting Issue organized and edited by R. J. H. Clark and P. Day,
  and N. S. Hush}}(2008)}]{Clark2008a}
\bibinfo{author}{\bibnamefont{{For a recent survey, see Mixed Valency: Papers
  of a Discussion Meeting Issue organized and edited by R. J. H. Clark and P.
  Day, and N. S. Hush}}}, \bibinfo{journal}{Phil. Trans. R. Soc.}
  \textbf{\bibinfo{volume}{A 366}}, \bibinfo{pages}{3} (\bibinfo{year}{2008}).

\bibitem[{\citenamefont{Robin and Day}(1967)}]{Robin1967}
\bibinfo{author}{\bibfnamefont{M.}~\bibnamefont{Robin}} \bibnamefont{and}
  \bibinfo{author}{\bibfnamefont{P.}~\bibnamefont{Day}}, \bibinfo{journal}{Adv.
  Inorg. Chem. Radiochem.} \textbf{\bibinfo{volume}{10}}, \bibinfo{pages}{247}
  (\bibinfo{year}{1967}).

\bibitem[{\citenamefont{Cox}(1987)}]{Cox1987}
\bibinfo{author}{\bibfnamefont{P.}~\bibnamefont{Cox}},
  \emph{\bibinfo{title}{The Electronic Structure and Chemistry of Solids}},
  \bibinfo{number}{Section 6.3.} (\bibinfo{publisher}{Oxford U.P., Oxford},
  \bibinfo{year}{1987}).

\bibitem[{\citenamefont{Varma}(1976)}]{Varma1976}
\bibinfo{author}{\bibfnamefont{C.~M.} \bibnamefont{Varma}},
  \bibinfo{journal}{Rev. Mod. Phys.} \textbf{\bibinfo{volume}{48}},
  \bibinfo{pages}{219} (\bibinfo{year}{1976}).

\bibitem[{\citenamefont{Gardner et~al.}(1972)\citenamefont{Gardner, Penfold,
  Smith, and Harris}}]{Gardner1972}
\bibinfo{author}{\bibfnamefont{W.~E.} \bibnamefont{Gardner}},
  \bibinfo{author}{\bibfnamefont{J.}~\bibnamefont{Penfold}},
  \bibinfo{author}{\bibfnamefont{T.~F.} \bibnamefont{Smith}}, \bibnamefont{and}
  \bibinfo{author}{\bibfnamefont{I.~R.} \bibnamefont{Harris}},
  \bibinfo{journal}{J. Phys. F} \textbf{\bibinfo{volume}{2}},
  \bibinfo{pages}{133} (\bibinfo{year}{1972}).

\bibitem[{\citenamefont{Batlogg et~al.}(1979)\citenamefont{Batlogg, Ott,
  Kaldis, Thoni, and Wachter}}]{Batlogg1979}
\bibinfo{author}{\bibfnamefont{B.}~\bibnamefont{Batlogg}},
  \bibinfo{author}{\bibfnamefont{H.~R.} \bibnamefont{Ott}},
  \bibinfo{author}{\bibfnamefont{E.}~\bibnamefont{Kaldis}},
  \bibinfo{author}{\bibfnamefont{W.}~\bibnamefont{Thoni}}, \bibnamefont{and}
  \bibinfo{author}{\bibfnamefont{P.}~\bibnamefont{Wachter}},
  \bibinfo{journal}{Phys. Rev. B} \textbf{\bibinfo{volume}{19}},
  \bibinfo{pages}{247} (\bibinfo{year}{1979}).

\bibitem[{\citenamefont{Adroja and Malik}(1991)}]{Adroja1991}
\bibinfo{author}{\bibfnamefont{D.~T.} \bibnamefont{Adroja}} \bibnamefont{and}
  \bibinfo{author}{\bibfnamefont{S.~K.} \bibnamefont{Malik}},
  \bibinfo{journal}{J. Magn. Magn. Mater.} \textbf{\bibinfo{volume}{100}},
  \bibinfo{pages}{126} (\bibinfo{year}{1991}).

\bibitem[{\citenamefont{Walz}(2002)}]{Walz2002}
\bibinfo{author}{\bibfnamefont{F.}~\bibnamefont{Walz}}, \bibinfo{journal}{J.
  Phys.: Condens. Matter} \textbf{\bibinfo{volume}{14}}, \bibinfo{pages}{R285}
  (\bibinfo{year}{2002}).

\bibitem[{\citenamefont{Batlogg et~al.}(1975)\citenamefont{Batlogg, Kaldis,
  Schlegel, and Wachter}}]{Batlogg1975}
\bibinfo{author}{\bibfnamefont{B.}~\bibnamefont{Batlogg}},
  \bibinfo{author}{\bibfnamefont{E.}~\bibnamefont{Kaldis}},
  \bibinfo{author}{\bibfnamefont{A.}~\bibnamefont{Schlegel}}, \bibnamefont{and}
  \bibinfo{author}{\bibfnamefont{P.}~\bibnamefont{Wachter}},
  \bibinfo{journal}{Phys. Rev. B} \textbf{\bibinfo{volume}{12}},
  \bibinfo{pages}{3940} (\bibinfo{year}{1975}).

\bibitem[{\citenamefont{Ohara et~al.}(2004)\citenamefont{Ohara, Sasaki,
  Konoike, Toyoda, Yamawaki, and Tanaka}}]{Ohara2004}
\bibinfo{author}{\bibfnamefont{H.}~\bibnamefont{Ohara}},
  \bibinfo{author}{\bibfnamefont{S.}~\bibnamefont{Sasaki}},
  \bibinfo{author}{\bibfnamefont{Y.}~\bibnamefont{Konoike}},
  \bibinfo{author}{\bibfnamefont{T.}~\bibnamefont{Toyoda}},
  \bibinfo{author}{\bibfnamefont{K.}~\bibnamefont{Yamawaki}}, \bibnamefont{and}
  \bibinfo{author}{\bibfnamefont{M.}~\bibnamefont{Tanaka}},
  \bibinfo{journal}{Physica B} \textbf{\bibinfo{volume}{350}},
  \bibinfo{pages}{353} (\bibinfo{year}{2004}).

\bibitem[{\citenamefont{Ravindran et~al.}(2008)\citenamefont{Ravindran, Vidya,
  Fjellvag, and Kjekshus}}]{Ravindran2008}
\bibinfo{author}{\bibfnamefont{P.}~\bibnamefont{Ravindran}},
  \bibinfo{author}{\bibfnamefont{R.}~\bibnamefont{Vidya}},
  \bibinfo{author}{\bibfnamefont{H.}~\bibnamefont{Fjellvag}}, \bibnamefont{and}
  \bibinfo{author}{\bibfnamefont{A.}~\bibnamefont{Kjekshus}},
  \bibinfo{journal}{Phys. Rev. B} \textbf{\bibinfo{volume}{77}},
  \bibinfo{pages}{134448} (\bibinfo{year}{2008}).

\bibitem[{\citenamefont{Trovarelli}(2002)}]{Trovarelli2002a}
\bibinfo{author}{\bibfnamefont{A.}~\bibnamefont{Trovarelli}},
  \emph{\bibinfo{title}{Catalysis by Ceria and Related Materials}}
  (\bibinfo{publisher}{Imperial College Press}, \bibinfo{year}{2002}).

\bibitem[{\citenamefont{Hoskins and Martin}(1995)}]{Hoskins1995}
\bibinfo{author}{\bibfnamefont{B.~F.} \bibnamefont{Hoskins}} \bibnamefont{and}
  \bibinfo{author}{\bibfnamefont{R.~L.} \bibnamefont{Martin}},
  \bibinfo{journal}{Australian Journal of Chemistry}
  \textbf{\bibinfo{volume}{48}}, \bibinfo{pages}{709} (\bibinfo{year}{1995}).

\bibitem[{\citenamefont{Skorodumova et~al.}(2002)\citenamefont{Skorodumova,
  Simak, Lundqvist, Abrikosov, and Johansson}}]{Skorodumova2002}
\bibinfo{author}{\bibfnamefont{N.~V.} \bibnamefont{Skorodumova}},
  \bibinfo{author}{\bibfnamefont{S.~I.} \bibnamefont{Simak}},
  \bibinfo{author}{\bibfnamefont{B.~I.} \bibnamefont{Lundqvist}},
  \bibinfo{author}{\bibfnamefont{I.~A.} \bibnamefont{Abrikosov}},
  \bibnamefont{and}
  \bibinfo{author}{\bibfnamefont{B.}~\bibnamefont{Johansson}},
  \bibinfo{journal}{Phys. Rev. Lett.} \textbf{\bibinfo{volume}{89}},
  \bibinfo{pages}{166601} (\bibinfo{year}{2002}).

\bibitem[{\citenamefont{Esch et~al.}(2005)\citenamefont{Esch, Fabris, Zhou,
  Montini, Africh, Fornasiero, Comelli, and Rosei}}]{Esch2005}
\bibinfo{author}{\bibfnamefont{F.}~\bibnamefont{Esch}},
  \bibinfo{author}{\bibfnamefont{S.}~\bibnamefont{Fabris}},
  \bibinfo{author}{\bibfnamefont{L.}~\bibnamefont{Zhou}},
  \bibinfo{author}{\bibfnamefont{T.}~\bibnamefont{Montini}},
  \bibinfo{author}{\bibfnamefont{C.}~\bibnamefont{Africh}},
  \bibinfo{author}{\bibfnamefont{P.}~\bibnamefont{Fornasiero}},
  \bibinfo{author}{\bibfnamefont{G.}~\bibnamefont{Comelli}}, \bibnamefont{and}
  \bibinfo{author}{\bibfnamefont{R.}~\bibnamefont{Rosei}},
  \bibinfo{journal}{Science} \textbf{\bibinfo{volume}{309}},
  \bibinfo{pages}{752} (\bibinfo{year}{2005}).

\bibitem[{\citenamefont{Brown}(1992)}]{Brown1992}
\bibinfo{author}{\bibfnamefont{I.~D.} \bibnamefont{Brown}},
  \bibinfo{journal}{Acta Crystallogr., Sect. B} \textbf{\bibinfo{volume}{48}},
  \bibinfo{pages}{553} (\bibinfo{year}{1992}).

\bibitem[{\citenamefont{Brown}(2002{\natexlab{a}})}]{Brown2002}
\bibinfo{author}{\bibfnamefont{I.~D.} \bibnamefont{Brown}},
  \emph{\bibinfo{title}{The Chemical Bond in Inorganic Chemistry: The Bond
  Valence Model}}, International Union of Crystallography
  (\bibinfo{publisher}{Oxford Science Publications},
  \bibinfo{year}{2002}{\natexlab{a}}).

\bibitem[{\citenamefont{Burdett}(1995)}]{Burdett1995a}
\bibinfo{author}{\bibfnamefont{J.~K.} \bibnamefont{Burdett}},
  \emph{\bibinfo{title}{Chemical Bonding in Solids}}
  (\bibinfo{publisher}{Oxford University Press}, \bibinfo{year}{1995}),
  chap.~\bibinfo{chapter}{6}.

\bibitem[{\citenamefont{Kotani et~al.}(1988)\citenamefont{Kotani, Jo, and
  Parlebas}}]{Kotani1988}
\bibinfo{author}{\bibfnamefont{A.}~\bibnamefont{Kotani}},
  \bibinfo{author}{\bibfnamefont{T.}~\bibnamefont{Jo}}, \bibnamefont{and}
  \bibinfo{author}{\bibfnamefont{J.~C.} \bibnamefont{Parlebas}},
  \bibinfo{journal}{Adv. Phys.} \textbf{\bibinfo{volume}{37}},
  \bibinfo{pages}{37} (\bibinfo{year}{1988}).

\bibitem[{\citenamefont{Lawrence et~al.}(1981)\citenamefont{Lawrence,
  Riseborough, and Parks}}]{Lawrence1981}
\bibinfo{author}{\bibfnamefont{J.~M.} \bibnamefont{Lawrence}},
  \bibinfo{author}{\bibfnamefont{P.~S.} \bibnamefont{Riseborough}},
  \bibnamefont{and} \bibinfo{author}{\bibfnamefont{R.~D.} \bibnamefont{Parks}},
  \bibinfo{journal}{Rep. Prog. Phys.} \textbf{\bibinfo{volume}{44}},
  \bibinfo{pages}{1} (\bibinfo{year}{1981}).

\bibitem[{\citenamefont{Kotani and Ogasawasara}(1992)}]{Kotani1992}
\bibinfo{author}{\bibfnamefont{A.}~\bibnamefont{Kotani}} \bibnamefont{and}
  \bibinfo{author}{\bibfnamefont{H.}~\bibnamefont{Ogasawasara}},
  \bibinfo{journal}{J. Electron. Spectrosc. Relat. Phenom.}
  \textbf{\bibinfo{volume}{60}}, \bibinfo{pages}{257} (\bibinfo{year}{1992}).

\bibitem[{\citenamefont{de~Groot and Kotani}(2008)}]{Groot2008}
\bibinfo{author}{\bibfnamefont{F.}~\bibnamefont{de~Groot}} \bibnamefont{and}
  \bibinfo{author}{\bibfnamefont{A.}~\bibnamefont{Kotani}},
  \emph{\bibinfo{title}{Core Level Spectroscopy of Solids}}, Advances in
  Condensed Matter Science (\bibinfo{publisher}{CRC Press},
  \bibinfo{year}{2008}).

\bibitem[{\citenamefont{Wuilloud et~al.}(1984)\citenamefont{Wuilloud, Delley,
  Schneider, and Baer}}]{Wuilloud1984}
\bibinfo{author}{\bibfnamefont{E.}~\bibnamefont{Wuilloud}},
  \bibinfo{author}{\bibfnamefont{B.}~\bibnamefont{Delley}},
  \bibinfo{author}{\bibfnamefont{W.-D.} \bibnamefont{Schneider}},
  \bibnamefont{and} \bibinfo{author}{\bibfnamefont{Y.}~\bibnamefont{Baer}},
  \bibinfo{journal}{Phys. Rev. Lett.} \textbf{\bibinfo{volume}{53}},
  \bibinfo{pages}{202} (\bibinfo{year}{1984}).

\bibitem[{\citenamefont{Jo and Kotani}(1985)}]{Jo1985}
\bibinfo{author}{\bibfnamefont{T.}~\bibnamefont{Jo}} \bibnamefont{and}
  \bibinfo{author}{\bibfnamefont{A.}~\bibnamefont{Kotani}},
  \bibinfo{journal}{Solid State Commun.} \textbf{\bibinfo{volume}{54}},
  \bibinfo{pages}{451} (\bibinfo{year}{1985}).

\bibitem[{\citenamefont{Kotani et~al.}(1985)\citenamefont{Kotani, Mizuta, Jo,
  and Parlebas}}]{Kotani1985}
\bibinfo{author}{\bibfnamefont{A.}~\bibnamefont{Kotani}},
  \bibinfo{author}{\bibfnamefont{H.}~\bibnamefont{Mizuta}},
  \bibinfo{author}{\bibfnamefont{T.}~\bibnamefont{Jo}}, \bibnamefont{and}
  \bibinfo{author}{\bibfnamefont{J.~C.} \bibnamefont{Parlebas}},
  \bibinfo{journal}{Solid State Commun.} \textbf{\bibinfo{volume}{53}},
  \bibinfo{pages}{805} (\bibinfo{year}{1985}).

\bibitem[{\citenamefont{Nakazawa et~al.}(1996)\citenamefont{Nakazawa, Tanaka,
  Uozumi, and Kotani}}]{Nakazawa1996}
\bibinfo{author}{\bibfnamefont{M.}~\bibnamefont{Nakazawa}},
  \bibinfo{author}{\bibfnamefont{S.}~\bibnamefont{Tanaka}},
  \bibinfo{author}{\bibfnamefont{T.}~\bibnamefont{Uozumi}}, \bibnamefont{and}
  \bibinfo{author}{\bibfnamefont{A.}~\bibnamefont{Kotani}},
  \bibinfo{journal}{J. Phys. Soc. Jpn.} \textbf{\bibinfo{volume}{65}},
  \bibinfo{pages}{2303} (\bibinfo{year}{1996}).

\bibitem[{\citenamefont{Fujimori}(1983{\natexlab{a}})}]{Fujimori1983}
\bibinfo{author}{\bibfnamefont{A.}~\bibnamefont{Fujimori}},
  \bibinfo{journal}{Phys. Rev. B} \textbf{\bibinfo{volume}{28}},
  \bibinfo{pages}{2281} (\bibinfo{year}{1983}{\natexlab{a}}).

\bibitem[{\citenamefont{Fujimori}(1983{\natexlab{b}})}]{Fujimori1983a}
\bibinfo{author}{\bibfnamefont{A.}~\bibnamefont{Fujimori}},
  \bibinfo{journal}{Phys. Rev. B} \textbf{\bibinfo{volume}{27}},
  \bibinfo{pages}{3992} (\bibinfo{year}{1983}{\natexlab{b}}).

\bibitem[{\citenamefont{Rao and Shripathi}(1997)}]{Rao1997}
\bibinfo{author}{\bibfnamefont{M.~V.~R.} \bibnamefont{Rao}} \bibnamefont{and}
  \bibinfo{author}{\bibfnamefont{T.}~\bibnamefont{Shripathi}},
  \bibinfo{journal}{J. Electron. Spectrosc. Relat. Phenom.}
  \textbf{\bibinfo{volume}{87}}, \bibinfo{pages}{121} (\bibinfo{year}{1997}).

\bibitem[{\citenamefont{Soldatov et~al.}(1994)\citenamefont{Soldatov,
  Ivanchenko, Longa, Kotani, Iwamoto, and Bianconi}}]{Soldatov1994}
\bibinfo{author}{\bibfnamefont{A.~V.} \bibnamefont{Soldatov}},
  \bibinfo{author}{\bibfnamefont{T.~S.} \bibnamefont{Ivanchenko}},
  \bibinfo{author}{\bibfnamefont{S.~D.} \bibnamefont{Longa}},
  \bibinfo{author}{\bibfnamefont{A.}~\bibnamefont{Kotani}},
  \bibinfo{author}{\bibfnamefont{Y.}~\bibnamefont{Iwamoto}}, \bibnamefont{and}
  \bibinfo{author}{\bibfnamefont{A.}~\bibnamefont{Bianconi}},
  \bibinfo{journal}{Phys. Rev. B} \textbf{\bibinfo{volume}{50}},
  \bibinfo{pages}{5074} (\bibinfo{year}{1994}).

\bibitem[{\citenamefont{Dexpert et~al.}(1987)\citenamefont{Dexpert, Karnatak,
  Esteva, Connerade, Gasgnier, Caro, and Albert}}]{Dexpert1987}
\bibinfo{author}{\bibfnamefont{H.}~\bibnamefont{Dexpert}},
  \bibinfo{author}{\bibfnamefont{R.~C.} \bibnamefont{Karnatak}},
  \bibinfo{author}{\bibfnamefont{J.~M.} \bibnamefont{Esteva}},
  \bibinfo{author}{\bibfnamefont{J.~P.} \bibnamefont{Connerade}},
  \bibinfo{author}{\bibfnamefont{M.}~\bibnamefont{Gasgnier}},
  \bibinfo{author}{\bibfnamefont{P.~E.} \bibnamefont{Caro}}, \bibnamefont{and}
  \bibinfo{author}{\bibfnamefont{L.}~\bibnamefont{Albert}},
  \bibinfo{journal}{Phys. Rev. B} \textbf{\bibinfo{volume}{36}},
  \bibinfo{pages}{1750} (\bibinfo{year}{1987}).

\bibitem[{\citenamefont{Bianconi et~al.}(1987)\citenamefont{Bianconi, Marcelli,
  Dexpert, Karnatak, Kotani, Jo, and Petiau}}]{Bianconi1987}
\bibinfo{author}{\bibfnamefont{A.}~\bibnamefont{Bianconi}},
  \bibinfo{author}{\bibfnamefont{A.}~\bibnamefont{Marcelli}},
  \bibinfo{author}{\bibfnamefont{H.}~\bibnamefont{Dexpert}},
  \bibinfo{author}{\bibfnamefont{R.}~\bibnamefont{Karnatak}},
  \bibinfo{author}{\bibfnamefont{A.}~\bibnamefont{Kotani}},
  \bibinfo{author}{\bibfnamefont{T.}~\bibnamefont{Jo}}, \bibnamefont{and}
  \bibinfo{author}{\bibfnamefont{J.}~\bibnamefont{Petiau}},
  \bibinfo{journal}{Phys. Rev. B} \textbf{\bibinfo{volume}{35}},
  \bibinfo{pages}{806} (\bibinfo{year}{1987}).

\bibitem[{\citenamefont{Karnatak et~al.}(1987)\citenamefont{Karnatak, Esteva,
  Dexpert, Gasgnier, Caro, and Albert}}]{Karnatak1987}
\bibinfo{author}{\bibfnamefont{R.~C.} \bibnamefont{Karnatak}},
  \bibinfo{author}{\bibfnamefont{J.~M.} \bibnamefont{Esteva}},
  \bibinfo{author}{\bibfnamefont{H.}~\bibnamefont{Dexpert}},
  \bibinfo{author}{\bibfnamefont{M.}~\bibnamefont{Gasgnier}},
  \bibinfo{author}{\bibfnamefont{P.~E.} \bibnamefont{Caro}}, \bibnamefont{and}
  \bibinfo{author}{\bibfnamefont{L.}~\bibnamefont{Albert}},
  \bibinfo{journal}{J. Magn. Magn. Mater.} \textbf{\bibinfo{volume}{63-64}},
  \bibinfo{pages}{518} (\bibinfo{year}{1987}).

\bibitem[{\citenamefont{Bianconi et~al.}(1982)\citenamefont{Bianconi, Campagna,
  and Stizza}}]{Bianconi1982}
\bibinfo{author}{\bibfnamefont{A.}~\bibnamefont{Bianconi}},
  \bibinfo{author}{\bibfnamefont{M.}~\bibnamefont{Campagna}}, \bibnamefont{and}
  \bibinfo{author}{\bibfnamefont{S.}~\bibnamefont{Stizza}},
  \bibinfo{journal}{Phys. Rev. B} \textbf{\bibinfo{volume}{25}},
  \bibinfo{pages}{2477} (\bibinfo{year}{1982}).

\bibitem[{\citenamefont{Finkelstein et~al.}(1992)\citenamefont{Finkelstein,
  Postnikov, Efremova, and Kurmaev}}]{Finkelstein1992}
\bibinfo{author}{\bibfnamefont{L.~D.} \bibnamefont{Finkelstein}},
  \bibinfo{author}{\bibfnamefont{A.~V.} \bibnamefont{Postnikov}},
  \bibinfo{author}{\bibfnamefont{N.~N.} \bibnamefont{Efremova}},
  \bibnamefont{and} \bibinfo{author}{\bibfnamefont{E.~Z.}
  \bibnamefont{Kurmaev}}, \bibinfo{journal}{Mater. Lett.}
  \textbf{\bibinfo{volume}{14}}, \bibinfo{pages}{115} (\bibinfo{year}{1992}).

\bibitem[{\citenamefont{Krill et~al.}(1981)\citenamefont{Krill, Kappler, Meyer,
  Abadli, and Ravet}}]{Krill1981}
\bibinfo{author}{\bibfnamefont{G.}~\bibnamefont{Krill}},
  \bibinfo{author}{\bibfnamefont{J.~P.} \bibnamefont{Kappler}},
  \bibinfo{author}{\bibfnamefont{A.}~\bibnamefont{Meyer}},
  \bibinfo{author}{\bibfnamefont{L.}~\bibnamefont{Abadli}}, \bibnamefont{and}
  \bibinfo{author}{\bibfnamefont{M.~F.} \bibnamefont{Ravet}},
  \bibinfo{journal}{J. Phys. F} \textbf{\bibinfo{volume}{11}},
  \bibinfo{pages}{1713} (\bibinfo{year}{1981}).

\bibitem[{\citenamefont{Kotani et~al.}(1987)\citenamefont{Kotani, Okada, and
  Jo}}]{Kotani1987}
\bibinfo{author}{\bibfnamefont{A.}~\bibnamefont{Kotani}},
  \bibinfo{author}{\bibfnamefont{M.}~\bibnamefont{Okada}}, \bibnamefont{and}
  \bibinfo{author}{\bibfnamefont{T.}~\bibnamefont{Jo}}, \bibinfo{journal}{J.
  Phys. Soc. Jpn.} \textbf{\bibinfo{volume}{56}}, \bibinfo{pages}{798}
  (\bibinfo{year}{1987}).

\bibitem[{\citenamefont{Haensel et~al.}(1970)\citenamefont{Haensel, Rabe, and
  Sonntag}}]{Haensel1970}
\bibinfo{author}{\bibfnamefont{R.}~\bibnamefont{Haensel}},
  \bibinfo{author}{\bibfnamefont{P.}~\bibnamefont{Rabe}}, \bibnamefont{and}
  \bibinfo{author}{\bibfnamefont{B.}~\bibnamefont{Sonntag}},
  \bibinfo{journal}{Solid State Commun.} \textbf{\bibinfo{volume}{8}},
  \bibinfo{pages}{1845} (\bibinfo{year}{1970}).

\bibitem[{\citenamefont{Hanyuu et~al.}(1985)\citenamefont{Hanyuu, Ishii,
  Yanagihara, Kamada, Miyahara, Kato, Naito, Suzuki, and Ishii}}]{Hanyuu1985}
\bibinfo{author}{\bibfnamefont{T.}~\bibnamefont{Hanyuu}},
  \bibinfo{author}{\bibfnamefont{H.}~\bibnamefont{Ishii}},
  \bibinfo{author}{\bibfnamefont{M.}~\bibnamefont{Yanagihara}},
  \bibinfo{author}{\bibfnamefont{T.}~\bibnamefont{Kamada}},
  \bibinfo{author}{\bibfnamefont{T.}~\bibnamefont{Miyahara}},
  \bibinfo{author}{\bibfnamefont{H.}~\bibnamefont{Kato}},
  \bibinfo{author}{\bibfnamefont{K.}~\bibnamefont{Naito}},
  \bibinfo{author}{\bibfnamefont{S.}~\bibnamefont{Suzuki}}, \bibnamefont{and}
  \bibinfo{author}{\bibfnamefont{T.}~\bibnamefont{Ishii}},
  \bibinfo{journal}{Solid State Commun.} \textbf{\bibinfo{volume}{56}},
  \bibinfo{pages}{381} (\bibinfo{year}{1985}).

\bibitem[{\citenamefont{Miyahara et~al.}(1987)\citenamefont{Miyahara, Fujimori,
  Koide, Sato, Shin, M.~Ishigame, and Komatsubara}}]{Miyahara1987}
\bibinfo{author}{\bibfnamefont{T.}~\bibnamefont{Miyahara}},
  \bibinfo{author}{\bibfnamefont{A.}~\bibnamefont{Fujimori}},
  \bibinfo{author}{\bibfnamefont{T.}~\bibnamefont{Koide}},
  \bibinfo{author}{\bibfnamefont{S.}~\bibnamefont{Sato}},
  \bibinfo{author}{\bibfnamefont{S.}~\bibnamefont{Shin}},
  \bibinfo{author}{\bibfnamefont{Y.~O.} \bibnamefont{M.~Ishigame}},
  \bibnamefont{and}
  \bibinfo{author}{\bibfnamefont{T.}~\bibnamefont{Komatsubara}},
  \bibinfo{journal}{J. Phys. Soc. Jpn.} \textbf{\bibinfo{volume}{56}},
  \bibinfo{pages}{3689} (\bibinfo{year}{1987}).

\bibitem[{\citenamefont{Matsumoto et~al.}(1994)\citenamefont{Matsumoto, Soda,
  Ichikawa, Tanaka, Taguchi, Jouda, Aita, Tezuka, and Shin}}]{Matsumoto1994}
\bibinfo{author}{\bibfnamefont{M.}~\bibnamefont{Matsumoto}},
  \bibinfo{author}{\bibfnamefont{K.}~\bibnamefont{Soda}},
  \bibinfo{author}{\bibfnamefont{K.}~\bibnamefont{Ichikawa}},
  \bibinfo{author}{\bibfnamefont{S.}~\bibnamefont{Tanaka}},
  \bibinfo{author}{\bibfnamefont{Y.}~\bibnamefont{Taguchi}},
  \bibinfo{author}{\bibfnamefont{K.}~\bibnamefont{Jouda}},
  \bibinfo{author}{\bibfnamefont{O.}~\bibnamefont{Aita}},
  \bibinfo{author}{\bibfnamefont{Y.}~\bibnamefont{Tezuka}}, \bibnamefont{and}
  \bibinfo{author}{\bibfnamefont{S.}~\bibnamefont{Shin}},
  \bibinfo{journal}{Phys. Rev. B} \textbf{\bibinfo{volume}{50}},
  \bibinfo{pages}{11340} (\bibinfo{year}{1994}).

\bibitem[{\citenamefont{Butorin et~al.}(1996)\citenamefont{Butorin, Mancini,
  Guo, Wassdahl, Nordgren, Nakazawa, Tanaka, Uozomi, Kotani, Ma
  et~al.}}]{Butorin1996}
\bibinfo{author}{\bibfnamefont{S.~M.} \bibnamefont{Butorin}},
  \bibinfo{author}{\bibfnamefont{D.~C.} \bibnamefont{Mancini}},
  \bibinfo{author}{\bibfnamefont{J.~H.} \bibnamefont{Guo}},
  \bibinfo{author}{\bibfnamefont{N.}~\bibnamefont{Wassdahl}},
  \bibinfo{author}{\bibfnamefont{J.}~\bibnamefont{Nordgren}},
  \bibinfo{author}{\bibfnamefont{M.}~\bibnamefont{Nakazawa}},
  \bibinfo{author}{\bibfnamefont{S.}~\bibnamefont{Tanaka}},
  \bibinfo{author}{\bibfnamefont{T.}~\bibnamefont{Uozomi}},
  \bibinfo{author}{\bibfnamefont{A.}~\bibnamefont{Kotani}},
  \bibinfo{author}{\bibfnamefont{Y.}~\bibnamefont{Ma}}, \bibnamefont{et~al.},
  \bibinfo{journal}{Phys. Rev. Lett.} \textbf{\bibinfo{volume}{77}},
  \bibinfo{pages}{574} (\bibinfo{year}{1996}).

\bibitem[{\citenamefont{Sham et~al.}(2005)\citenamefont{Sham, Gordon, and
  Heald}}]{Sham2005a}
\bibinfo{author}{\bibfnamefont{T.~K.} \bibnamefont{Sham}},
  \bibinfo{author}{\bibfnamefont{R.~A.} \bibnamefont{Gordon}},
  \bibnamefont{and} \bibinfo{author}{\bibfnamefont{S.~M.} \bibnamefont{Heald}},
  \bibinfo{journal}{Phys. Rev. B} \textbf{\bibinfo{volume}{72}},
  \bibinfo{pages}{035113} (\bibinfo{year}{2005}).

\bibitem[{\citenamefont{Allen}(1985)}]{Allen1985}
\bibinfo{author}{\bibfnamefont{J.~W.} \bibnamefont{Allen}},
  \bibinfo{journal}{J. Magn. Magn. Mater.} \textbf{\bibinfo{volume}{47-48}},
  \bibinfo{pages}{168} (\bibinfo{year}{1985}).

\bibitem[{\citenamefont{Orchard and Thornton}(1977)}]{Orchard1977}
\bibinfo{author}{\bibfnamefont{A.~F.} \bibnamefont{Orchard}} \bibnamefont{and}
  \bibinfo{author}{\bibfnamefont{G.}~\bibnamefont{Thornton}},
  \bibinfo{journal}{J. Electron. Spectrosc. Relat. Phenom.}
  \textbf{\bibinfo{volume}{10}}, \bibinfo{pages}{1} (\bibinfo{year}{1977}).

\bibitem[{\citenamefont{Ryzhkov et~al.}(1985)\citenamefont{Ryzhkov, Gubanov,
  Teterin, and Baev}}]{Ryzhkov1985}
\bibinfo{author}{\bibfnamefont{M.~V.} \bibnamefont{Ryzhkov}},
  \bibinfo{author}{\bibfnamefont{V.~A.} \bibnamefont{Gubanov}},
  \bibinfo{author}{\bibfnamefont{Y.~A.} \bibnamefont{Teterin}},
  \bibnamefont{and} \bibinfo{author}{\bibfnamefont{A.~S.} \bibnamefont{Baev}},
  \bibinfo{journal}{Z. Phys. B} \textbf{\bibinfo{volume}{59}},
  \bibinfo{pages}{1} (\bibinfo{year}{1985}).

\bibitem[{\citenamefont{Nakano et~al.}(1987)\citenamefont{Nakano, Kotani, and
  Parlebas}}]{Nakano1987a}
\bibinfo{author}{\bibfnamefont{T.}~\bibnamefont{Nakano}},
  \bibinfo{author}{\bibfnamefont{A.}~\bibnamefont{Kotani}}, \bibnamefont{and}
  \bibinfo{author}{\bibfnamefont{J.~C.} \bibnamefont{Parlebas}},
  \bibinfo{journal}{J. Phys. Soc. Jpn.} \textbf{\bibinfo{volume}{56}},
  \bibinfo{pages}{2201} (\bibinfo{year}{1987}).

\bibitem[{\citenamefont{Jo and Kotani}(1988)}]{Jo1988}
\bibinfo{author}{\bibfnamefont{T.}~\bibnamefont{Jo}} \bibnamefont{and}
  \bibinfo{author}{\bibfnamefont{A.}~\bibnamefont{Kotani}},
  \bibinfo{journal}{Phys. Rev. B} \textbf{\bibinfo{volume}{38}},
  \bibinfo{pages}{830} (\bibinfo{year}{1988}).

\bibitem[{\citenamefont{Koelling et~al.}(1983)\citenamefont{Koelling, Boring,
  and Wood}}]{Koelling1983}
\bibinfo{author}{\bibfnamefont{D.~D.} \bibnamefont{Koelling}},
  \bibinfo{author}{\bibfnamefont{A.~M.} \bibnamefont{Boring}},
  \bibnamefont{and} \bibinfo{author}{\bibfnamefont{J.~H.} \bibnamefont{Wood}},
  \bibinfo{journal}{Solid State Commun.} \textbf{\bibinfo{volume}{47}},
  \bibinfo{pages}{227} (\bibinfo{year}{1983}).

\bibitem[{\citenamefont{Thornton and Dempsey}(1981)}]{Thornton1981}
\bibinfo{author}{\bibfnamefont{G.}~\bibnamefont{Thornton}} \bibnamefont{and}
  \bibinfo{author}{\bibfnamefont{M.~J.} \bibnamefont{Dempsey}},
  \bibinfo{journal}{Chem. Phys. Lett.} \textbf{\bibinfo{volume}{77}},
  \bibinfo{pages}{409} (\bibinfo{year}{1981}).

\bibitem[{\citenamefont{Hay et~al.}(2006)\citenamefont{Hay, Martin, Uddin, and
  Scuseria}}]{Hay2006}
\bibinfo{author}{\bibfnamefont{P.~J.} \bibnamefont{Hay}},
  \bibinfo{author}{\bibfnamefont{R.~L.} \bibnamefont{Martin}},
  \bibinfo{author}{\bibfnamefont{J.}~\bibnamefont{Uddin}}, \bibnamefont{and}
  \bibinfo{author}{\bibfnamefont{G.~E.} \bibnamefont{Scuseria}},
  \bibinfo{journal}{J. Chem. Phys.} \textbf{\bibinfo{volume}{125}},
  \bibinfo{pages}{034712} (\bibinfo{year}{2006}).

\bibitem[{\citenamefont{Andersson et~al.}(2007)\citenamefont{Andersson, Simak,
  Johansson, Abrikosov, and Skorodumova}}]{Andersson2007}
\bibinfo{author}{\bibfnamefont{D.~A.} \bibnamefont{Andersson}},
  \bibinfo{author}{\bibfnamefont{S.~I.} \bibnamefont{Simak}},
  \bibinfo{author}{\bibfnamefont{B.}~\bibnamefont{Johansson}},
  \bibinfo{author}{\bibfnamefont{I.~A.} \bibnamefont{Abrikosov}},
  \bibnamefont{and} \bibinfo{author}{\bibfnamefont{N.~V.}
  \bibnamefont{Skorodumova}}, \bibinfo{journal}{Phys. Rev. B}
  \textbf{\bibinfo{volume}{75}}, \bibinfo{pages}{035109}
  (\bibinfo{year}{2007}).

\bibitem[{\citenamefont{Fabris et~al.}(2005)\citenamefont{Fabris, de~Gironcoli,
  Baroni, Vicario, and Balducci}}]{Fabris2005a}
\bibinfo{author}{\bibfnamefont{S.}~\bibnamefont{Fabris}},
  \bibinfo{author}{\bibfnamefont{S.}~\bibnamefont{de~Gironcoli}},
  \bibinfo{author}{\bibfnamefont{S.}~\bibnamefont{Baroni}},
  \bibinfo{author}{\bibfnamefont{G.}~\bibnamefont{Vicario}}, \bibnamefont{and}
  \bibinfo{author}{\bibfnamefont{G.}~\bibnamefont{Balducci}},
  \bibinfo{journal}{Phys. Rev. B} \textbf{\bibinfo{volume}{71}},
  \bibinfo{pages}{041102(R)} (\bibinfo{year}{2005}).

\bibitem[{\citenamefont{Skorodumova et~al.}(2001)\citenamefont{Skorodumova,
  Ahuja, Simak, Abrikosov, Johansson, and Lundqvist}}]{Skorodumova2001}
\bibinfo{author}{\bibfnamefont{N.~V.} \bibnamefont{Skorodumova}},
  \bibinfo{author}{\bibfnamefont{R.}~\bibnamefont{Ahuja}},
  \bibinfo{author}{\bibfnamefont{S.~I.} \bibnamefont{Simak}},
  \bibinfo{author}{\bibfnamefont{I.~A.} \bibnamefont{Abrikosov}},
  \bibinfo{author}{\bibfnamefont{B.}~\bibnamefont{Johansson}},
  \bibnamefont{and} \bibinfo{author}{\bibfnamefont{B.~I.}
  \bibnamefont{Lundqvist}}, \bibinfo{journal}{Phys. Rev. B}
  \textbf{\bibinfo{volume}{64}}, \bibinfo{pages}{115108}
  (\bibinfo{year}{2001}).

\bibitem[{\citenamefont{Eriksson et~al.}(1991)\citenamefont{Eriksson, Albers,
  Boring, Fernando, Hao, and Cooper}}]{Eriksson1991}
\bibinfo{author}{\bibfnamefont{O.}~\bibnamefont{Eriksson}},
  \bibinfo{author}{\bibfnamefont{R.~C.} \bibnamefont{Albers}},
  \bibinfo{author}{\bibfnamefont{A.~M.} \bibnamefont{Boring}},
  \bibinfo{author}{\bibfnamefont{G.~W.} \bibnamefont{Fernando}},
  \bibinfo{author}{\bibfnamefont{Y.~G.} \bibnamefont{Hao}}, \bibnamefont{and}
  \bibinfo{author}{\bibfnamefont{B.~R.} \bibnamefont{Cooper}},
  \bibinfo{journal}{Phys. Rev. B} \textbf{\bibinfo{volume}{43}},
  \bibinfo{pages}{3137} (\bibinfo{year}{1991}).

\bibitem[{\citenamefont{Rueff et~al.}(2006)\citenamefont{Rueff, Itie, Hague,
  J.M, Delaunay, J.P, and Jaouen}}]{Rueff2006}
\bibinfo{author}{\bibfnamefont{J.}~\bibnamefont{Rueff}},
  \bibinfo{author}{\bibfnamefont{J.}~\bibnamefont{Itie}},
  \bibinfo{author}{\bibfnamefont{C.}~\bibnamefont{Hague}},
  \bibinfo{author}{\bibfnamefont{M.}~\bibnamefont{J.M}},
  \bibinfo{author}{\bibfnamefont{R.}~\bibnamefont{Delaunay}},
  \bibinfo{author}{\bibfnamefont{K.}~\bibnamefont{J.P}}, \bibnamefont{and}
  \bibinfo{author}{\bibfnamefont{N.}~\bibnamefont{Jaouen}},
  \bibinfo{journal}{Phys. Rev. Lett.} \textbf{\bibinfo{volume}{96}},
  \bibinfo{pages}{237403} (\bibinfo{year}{2006}).

\bibitem[{\citenamefont{Mohri}(2003)}]{Mohri2003}
\bibinfo{author}{\bibfnamefont{F.}~\bibnamefont{Mohri}}, \bibinfo{journal}{Acta
  Crystallogr., Sect. B} \textbf{\bibinfo{volume}{59}}, \bibinfo{pages}{190}
  (\bibinfo{year}{2003}).

\bibitem[{\citenamefont{Brese and O'Keeffe}(1991)}]{Brese1991}
\bibinfo{author}{\bibfnamefont{N.~E.} \bibnamefont{Brese}} \bibnamefont{and}
  \bibinfo{author}{\bibfnamefont{M.}~\bibnamefont{O'Keeffe}},
  \bibinfo{journal}{Acta. Crystallogr., Sect. B} \textbf{\bibinfo{volume}{47}},
  \bibinfo{pages}{192} (\bibinfo{year}{1991}).

\bibitem[{\citenamefont{Brown}(2002{\natexlab{b}})}]{Brown2002a}
\bibinfo{author}{\bibfnamefont{I.}~\bibnamefont{Brown}},
  \emph{\bibinfo{title}{The Chemical Bond in Inorganic Chemistry: The Bond
  Valence Model}} (\bibinfo{publisher}{Oxford Science Publications},
  \bibinfo{year}{2002}{\natexlab{b}}), chap.~\bibinfo{chapter}{4},
  p.~\bibinfo{pages}{43}.

\bibitem[{\citenamefont{Altermatt and Brown}(1985)}]{Altermatt1985}
\bibinfo{author}{\bibfnamefont{D.}~\bibnamefont{Altermatt}} \bibnamefont{and}
  \bibinfo{author}{\bibfnamefont{I.~D.} \bibnamefont{Brown}},
  \bibinfo{journal}{Acta Crystallogr.} \textbf{\bibinfo{volume}{B41}},
  \bibinfo{pages}{240} (\bibinfo{year}{1985}).

\bibitem[{\citenamefont{Roulhac and Palenik}(2003)}]{Roulhac2003}
\bibinfo{author}{\bibfnamefont{P.~L.} \bibnamefont{Roulhac}} \bibnamefont{and}
  \bibinfo{author}{\bibfnamefont{G.~J.} \bibnamefont{Palenik}},
  \bibinfo{journal}{Inorg. Chem.} \textbf{\bibinfo{volume}{42}},
  \bibinfo{pages}{118} (\bibinfo{year}{2003}).

\bibitem[{\citenamefont{Trzesowska et~al.}(2004)\citenamefont{Trzesowska,
  Kruszynski, and Bartczak}}]{Trzesowska2004}
\bibinfo{author}{\bibfnamefont{A.}~\bibnamefont{Trzesowska}},
  \bibinfo{author}{\bibfnamefont{R.}~\bibnamefont{Kruszynski}},
  \bibnamefont{and} \bibinfo{author}{\bibfnamefont{T.}~\bibnamefont{Bartczak}},
  \bibinfo{journal}{Acta Crystallogr., Sect. B} \textbf{\bibinfo{volume}{60}},
  \bibinfo{pages}{174} (\bibinfo{year}{2004}).

\bibitem[{\citenamefont{Zocchi}(2007)}]{Zocchi2007}
\bibinfo{author}{\bibfnamefont{F.}~\bibnamefont{Zocchi}}, \bibinfo{journal}{J.
  Mol. Struc. - THEOCHEM} \textbf{\bibinfo{volume}{805}}, \bibinfo{pages}{73}
  (\bibinfo{year}{2007}).

\bibitem[{\citenamefont{Zocchi}(2002)}]{Zocchi2002}
\bibinfo{author}{\bibfnamefont{F.}~\bibnamefont{Zocchi}},
  \bibinfo{journal}{Solid State Sci.} \textbf{\bibinfo{volume}{4}},
  \bibinfo{pages}{149} (\bibinfo{year}{2002}).

\bibitem[{\citenamefont{CrystalMaker$^{\textregistered}$}(2008)}]{CrystalMaker}
\bibinfo{author}{\bibnamefont{CrystalMaker$^{\textregistered}$}},
  \emph{\bibinfo{title}{:{A} crystal and molecular structures program for {Mac}
  and {Windows}.}} (\bibinfo{year}{2008}).

\bibitem[{\citenamefont{Eyring}(1979)}]{Eyring1979}
\bibinfo{author}{\bibfnamefont{L.}~\bibnamefont{Eyring}}, in
  \emph{\bibinfo{booktitle}{Handbook on the physics and chemistry of rare
  earths}}, edited by \bibinfo{editor}{\bibfnamefont{K.~A.}
  \bibnamefont{Gschneider}} \bibnamefont{and}
  \bibinfo{editor}{\bibfnamefont{L.}~\bibnamefont{Eyring}}
  (\bibinfo{year}{1979}), vol.~\bibinfo{volume}{3}, p. \bibinfo{pages}{Chap.
  27}.

\bibitem[{\citenamefont{Tsunekawa et~al.}(1999)\citenamefont{Tsunekawa,
  Sivamohan, Ito, Kasuya, and Fukuda}}]{Tsunekawa1999}
\bibinfo{author}{\bibfnamefont{S.}~\bibnamefont{Tsunekawa}},
  \bibinfo{author}{\bibfnamefont{R.}~\bibnamefont{Sivamohan}},
  \bibinfo{author}{\bibfnamefont{S.}~\bibnamefont{Ito}},
  \bibinfo{author}{\bibfnamefont{A.}~\bibnamefont{Kasuya}}, \bibnamefont{and}
  \bibinfo{author}{\bibfnamefont{T.}~\bibnamefont{Fukuda}},
  \bibinfo{journal}{Nanostruct. Mater.} \textbf{\bibinfo{volume}{11}},
  \bibinfo{pages}{141} (\bibinfo{year}{1999}).

\bibitem[{\citenamefont{Sorensen}(1976)}]{Sorensen1976}
\bibinfo{author}{\bibfnamefont{O.~T.} \bibnamefont{Sorensen}},
  \bibinfo{journal}{J. Solid State Chem.} \textbf{\bibinfo{volume}{18}},
  \bibinfo{pages}{217} (\bibinfo{year}{1976}).

\bibitem[{\citenamefont{Barnighausen and Schiller}(1985)}]{Barnighausen1985}
\bibinfo{author}{\bibfnamefont{H.}~\bibnamefont{Barnighausen}}
  \bibnamefont{and} \bibinfo{author}{\bibfnamefont{G.}~\bibnamefont{Schiller}},
  \bibinfo{journal}{J. Less-Common Met.} \textbf{\bibinfo{volume}{110}},
  \bibinfo{pages}{385} (\bibinfo{year}{1985}).

\bibitem[{\citenamefont{Wyckoff}(1963)}]{Wyckoff1964}
\bibinfo{author}{\bibfnamefont{R.~W.~G.} \bibnamefont{Wyckoff}},
  \emph{\bibinfo{title}{Crystal Structures}}, vol.~\bibinfo{volume}{2}
  (\bibinfo{publisher}{Interscience}, \bibinfo{year}{1963}),
  \bibinfo{edition}{2nd} ed.

\bibitem[{\citenamefont{Villars and Calvert}(1991)}]{Villars1991}
\bibinfo{author}{\bibfnamefont{P.}~\bibnamefont{Villars}} \bibnamefont{and}
  \bibinfo{author}{\bibfnamefont{L.~D.} \bibnamefont{Calvert}},
  \emph{\bibinfo{title}{Pearson's Handbook of Crystallographic Data for
  Intermetallic Phases}}, vol.~\bibinfo{volume}{2} (\bibinfo{publisher}{ASM
  International}, \bibinfo{year}{1991}).

\bibitem[{\citenamefont{Kummerle and Heger}(1999)}]{Kummerle1999}
\bibinfo{author}{\bibfnamefont{E.~A.} \bibnamefont{Kummerle}} \bibnamefont{and}
  \bibinfo{author}{\bibfnamefont{G.}~\bibnamefont{Heger}}, \bibinfo{journal}{J.
  Solid State Chem.} \textbf{\bibinfo{volume}{147}}, \bibinfo{pages}{485}
  (\bibinfo{year}{1999}).

\bibitem[{\citenamefont{Zinkevich et~al.}(2006)\citenamefont{Zinkevich,
  Djurovic, and Aldinger}}]{Zinkevich2006}
\bibinfo{author}{\bibfnamefont{M.}~\bibnamefont{Zinkevich}},
  \bibinfo{author}{\bibfnamefont{D.}~\bibnamefont{Djurovic}}, \bibnamefont{and}
  \bibinfo{author}{\bibfnamefont{F.}~\bibnamefont{Aldinger}},
  \bibinfo{journal}{Solid State Ionics} \textbf{\bibinfo{volume}{177}},
  \bibinfo{pages}{989} (\bibinfo{year}{2006}).

\bibitem[{\citenamefont{Bartram}(1966)}]{Bartram1966}
\bibinfo{author}{\bibfnamefont{S.~F.} \bibnamefont{Bartram}},
  \bibinfo{journal}{Inorg. Chem.} \textbf{\bibinfo{volume}{5}},
  \bibinfo{pages}{749} (\bibinfo{year}{1966}).

\bibitem[{\citenamefont{Ray and Cox}(1975)}]{Ray1975}
\bibinfo{author}{\bibfnamefont{S.~P.} \bibnamefont{Ray}} \bibnamefont{and}
  \bibinfo{author}{\bibfnamefont{D.~E.} \bibnamefont{Cox}},
  \bibinfo{journal}{J. Solid State Chem.} \textbf{\bibinfo{volume}{15}},
  \bibinfo{pages}{333} (\bibinfo{year}{1975}).

\bibitem[{\citenamefont{Zhang et~al.}(1993)\citenamefont{Zhang, Dreele, and
  Eyring}}]{Zhang1993}
\bibinfo{author}{\bibfnamefont{J.}~\bibnamefont{Zhang}},
  \bibinfo{author}{\bibfnamefont{R.~V.} \bibnamefont{Dreele}},
  \bibnamefont{and} \bibinfo{author}{\bibfnamefont{L.}~\bibnamefont{Eyring}},
  \bibinfo{journal}{J. Solid State Chem.} \textbf{\bibinfo{volume}{104}},
  \bibinfo{pages}{21} (\bibinfo{year}{1993}).

\bibitem[{\citenamefont{Zhang et~al.}(1996)\citenamefont{Zhang, Von~Dreele, and
  Eyring}}]{Zhang1996}
\bibinfo{author}{\bibfnamefont{J.}~\bibnamefont{Zhang}},
  \bibinfo{author}{\bibfnamefont{R.~B.} \bibnamefont{Von~Dreele}},
  \bibnamefont{and} \bibinfo{author}{\bibfnamefont{L.}~\bibnamefont{Eyring}},
  \bibinfo{journal}{J. Solid State Chem.} \textbf{\bibinfo{volume}{122}},
  \bibinfo{pages}{53} (\bibinfo{year}{1996}).

\bibitem[{\citenamefont{Taylor}(1984)}]{Taylor1984}
\bibinfo{author}{\bibfnamefont{D.}~\bibnamefont{Taylor}},
  \bibinfo{journal}{Trans. J. Br. Ceram. Soc.} \textbf{\bibinfo{volume}{83}},
  \bibinfo{pages}{32} (\bibinfo{year}{1984}).

\bibitem[{\citenamefont{Brown}(1981)}]{Brown1981}
\bibinfo{author}{\bibfnamefont{I.~D.} \bibnamefont{Brown}},
  \emph{\bibinfo{title}{The bond-valence method: an empirical approach to
  chemical structure and bonding}} (\bibinfo{publisher}{Academic Press, New
  York}, \bibinfo{year}{1981}), vol.~\bibinfo{volume}{II},
  chap.~\bibinfo{chapter}{1}, pp. \bibinfo{pages}{1--30}.

\bibitem[{\citenamefont{Locock and Burns}(2004)}]{Locock2004}
\bibinfo{author}{\bibfnamefont{A.~J.} \bibnamefont{Locock}} \bibnamefont{and}
  \bibinfo{author}{\bibfnamefont{P.~C.} \bibnamefont{Burns}},
  \bibinfo{journal}{Z. Kristallogr.} \textbf{\bibinfo{volume}{219}},
  \bibinfo{pages}{259} (\bibinfo{year}{2004}).

\bibitem[{\citenamefont{Krivovichev and Brown}(2001)}]{Krivovichev2001}
\bibinfo{author}{\bibfnamefont{S.~V.} \bibnamefont{Krivovichev}}
  \bibnamefont{and} \bibinfo{author}{\bibfnamefont{I.~D.} \bibnamefont{Brown}},
  \bibinfo{journal}{Z. Kristallogr.} \textbf{\bibinfo{volume}{216}},
  \bibinfo{pages}{245} (\bibinfo{year}{2001}).

\bibitem[{\citenamefont{Liebau and Wang}(2005)}]{Liebau2005}
\bibinfo{author}{\bibfnamefont{F.}~\bibnamefont{Liebau}} \bibnamefont{and}
  \bibinfo{author}{\bibfnamefont{X.}~\bibnamefont{Wang}}, \bibinfo{journal}{Z.
  Kristallogr.} \textbf{\bibinfo{volume}{220}}, \bibinfo{pages}{589}
  (\bibinfo{year}{2005}).

\bibitem[{\citenamefont{Wu et~al.}(2001)\citenamefont{Wu, Benfield, Li, Yang,
  Grandjean, Li, and Zhu}}]{Wu2001}
\bibinfo{author}{\bibfnamefont{Z.}~\bibnamefont{Wu}},
  \bibinfo{author}{\bibfnamefont{R.~E.} \bibnamefont{Benfield}},
  \bibinfo{author}{\bibfnamefont{L.~G.~H.} \bibnamefont{Li}},
  \bibinfo{author}{\bibfnamefont{Q.}~\bibnamefont{Yang}},
  \bibinfo{author}{\bibfnamefont{D.}~\bibnamefont{Grandjean}},
  \bibinfo{author}{\bibfnamefont{Q.}~\bibnamefont{Li}}, \bibnamefont{and}
  \bibinfo{author}{\bibfnamefont{H.}~\bibnamefont{Zhu}}, \bibinfo{journal}{J.
  Phys.: Condens. Matter} \textbf{\bibinfo{volume}{13}}, \bibinfo{pages}{5269}
  (\bibinfo{year}{2001}).

\bibitem[{\citenamefont{Marabelli and Wachter}(1987)}]{Marabelli1987}
\bibinfo{author}{\bibfnamefont{F.}~\bibnamefont{Marabelli}} \bibnamefont{and}
  \bibinfo{author}{\bibfnamefont{P.}~\bibnamefont{Wachter}},
  \bibinfo{journal}{Phys. Rev. B} \textbf{\bibinfo{volume}{36}},
  \bibinfo{pages}{1238} (\bibinfo{year}{1987}).

\bibitem[{\citenamefont{Castleton et~al.}(2007)\citenamefont{Castleton,
  Kullgren, and Hermansson}}]{Castleton2007}
\bibinfo{author}{\bibfnamefont{C.~W.~M.} \bibnamefont{Castleton}},
  \bibinfo{author}{\bibfnamefont{J.}~\bibnamefont{Kullgren}}, \bibnamefont{and}
  \bibinfo{author}{\bibfnamefont{K.}~\bibnamefont{Hermansson}},
  \bibinfo{journal}{J. Chem. Phys.} \textbf{\bibinfo{volume}{127}},
  \bibinfo{pages}{244704} (\bibinfo{year}{2007}).

\bibitem[{\citenamefont{Kotani et~al.}(1989)\citenamefont{Kotani, Ogasawara,
  Okada, Thole, and Sawatzky}}]{Kotani1989}
\bibinfo{author}{\bibfnamefont{A.}~\bibnamefont{Kotani}},
  \bibinfo{author}{\bibfnamefont{H.}~\bibnamefont{Ogasawara}},
  \bibinfo{author}{\bibfnamefont{K.}~\bibnamefont{Okada}},
  \bibinfo{author}{\bibfnamefont{B.~T.} \bibnamefont{Thole}}, \bibnamefont{and}
  \bibinfo{author}{\bibfnamefont{G.~A.} \bibnamefont{Sawatzky}},
  \bibinfo{journal}{Phys. Rev. B} \textbf{\bibinfo{volume}{40}},
  \bibinfo{pages}{65} (\bibinfo{year}{1989}).

\bibitem[{\citenamefont{Arai et~al.}(2004)\citenamefont{Arai, Muto, Murai,
  Sasaki, Ukyo, Kuroda, and Saka}}]{Arai2004}
\bibinfo{author}{\bibfnamefont{S.}~\bibnamefont{Arai}},
  \bibinfo{author}{\bibfnamefont{S.}~\bibnamefont{Muto}},
  \bibinfo{author}{\bibfnamefont{J.}~\bibnamefont{Murai}},
  \bibinfo{author}{\bibfnamefont{T.}~\bibnamefont{Sasaki}},
  \bibinfo{author}{\bibfnamefont{Y.}~\bibnamefont{Ukyo}},
  \bibinfo{author}{\bibfnamefont{K.}~\bibnamefont{Kuroda}}, \bibnamefont{and}
  \bibinfo{author}{\bibfnamefont{H.}~\bibnamefont{Saka}},
  \bibinfo{journal}{Mater. Trans.} \textbf{\bibinfo{volume}{45}},
  \bibinfo{pages}{2951} (\bibinfo{year}{2004}).

\bibitem[{\citenamefont{Shannon}(1976)}]{Shannon1976}
\bibinfo{author}{\bibfnamefont{R.~D.} \bibnamefont{Shannon}},
  \bibinfo{journal}{Acta. Crystallogr., Sect. A} \textbf{\bibinfo{volume}{32}},
  \bibinfo{pages}{751} (\bibinfo{year}{1976}).

\bibitem[{\citenamefont{Guillaume et~al.}(1993)\citenamefont{Guillaume,
  Allenspach, Mesot, Roessli, Staub, Fischer, and Furrer}}]{Guillaume1993}
\bibinfo{author}{\bibfnamefont{M.}~\bibnamefont{Guillaume}},
  \bibinfo{author}{\bibfnamefont{P.}~\bibnamefont{Allenspach}},
  \bibinfo{author}{\bibfnamefont{J.}~\bibnamefont{Mesot}},
  \bibinfo{author}{\bibfnamefont{B.}~\bibnamefont{Roessli}},
  \bibinfo{author}{\bibfnamefont{U.}~\bibnamefont{Staub}},
  \bibinfo{author}{\bibfnamefont{P.}~\bibnamefont{Fischer}}, \bibnamefont{and}
  \bibinfo{author}{\bibfnamefont{A.}~\bibnamefont{Furrer}},
  \bibinfo{journal}{Z. Phys. B} \textbf{\bibinfo{volume}{90}},
  \bibinfo{pages}{13} (\bibinfo{year}{1993}).

\end{thebibliography}

\appendix

\section{Bond Lengths in the Various Cerium Coordination Polyhedra of the
Cerium Oxides}

The calculated Ce-O bond distances in the various Ce coordination polyhedra of
the Ce oxide crystals are listed in Table \ref{lengths}.%

\begin{table*}[ptb] \centering
\caption{Bond length data used in the bond valence calculations of the Ce
oxides. Crystallographic data was obtained from the references as cited in the table and the Ce-O bond lengths were calculated from that data in CrystalMaker$^{\textregistered}$. \cite{CrystalMaker} The O atoms, O($i$) ($i = 1-8$) in each polyhedron may be equivalent or distinct O sites in a given crystal. All bond lengths are in \text{\AA}. Estimated errors in bond lengths are approximately ${±0.01}$ ${\text{\AA} }$. }
\begin{tabular*}{\textwidth}{@{\extracolsep{\fill}} *{11}c }
\hline\hline
Oxide & Ref. & Ce Site & \multicolumn{8}{c}{Bond Lengths in Ce
Coordination Polyhedron} \\\hline
&  &  & O(1) & O(2) & O(3) & O(4) & O(5) & O(6) & O(7) & O(8)\\\hline
A-Ce$_{2}$O$_{3}$ & \onlinecite{Barnighausen1985,Wyckoff1964} & Ce & $2.339$ &
$2.339$ & $2.339$ & $2.434$ & $2.694$ & $2.694$ & $2.694$ & $-$\\
C-Ce$_{2}$O$_{3}$ & \onlinecite{Wyckoff1964,Villars1991,Kummerle1999} &
Ce(1) & $2.416$ & $2.416$ & $2.416$ & $2.416$ & $2.416$ & $2.416$ & $-$ &
$-$\\
&  & Ce(2) & $2.375$ & $2.375$ & $2.403$ & $2.403$ & $2.469$ & $2.469$ & $-$ &
$-$\\
Ce$_{3}$O$_{5}$ & \onlinecite{Kummerle1999,Zinkevich2006,Wyckoff1964} &
Ce(1) & $2.393$ & $2.393$ & $2.393$ & $2.393$ & $2.393$ & $2.393$ & $2.600$ &
$2.600$\\
&  & Ce(2) & $2.352$ & $2.352$ & $2.380$ & $2.380$ & $2.445$ & $2.445$ &
$2.563$ & $2.563$\\
Ce$_{7}$O$_{12}$ & \onlinecite{Bartram1966,Ray1975} & Ce(1) & $2.249$ &
$2.249$ & $2.249$ & $2.249$ & $2.249$ & $2.249$ & $-$ & $-$\\
&  & Ce(2) & $2.405$ & $2.392$ & $2.652$ & $2.307$ & $2.367$ & $2.380$ &
$2.478$ & $-$\\
Ce$_{11}$O$_{20}$ & \onlinecite{Kummerle1999,Zhang1993} & Ce(1) & $2.489$ &
$2.474$ & $2.379$ & $2.489$ & $2.474$ & $2.379$ & $2.701$ & $2.701$\\
&  & Ce(2) & $2.423$ & $2.407$ & $2.446$ & $2.624$ & $2.589$ & $2.444$ &
$2.539$ & $2.597$\\
&  & Ce(3) & $2.311$ & $2.405$ & $2.263$ & $2.333$ & $2.447$ & $2.385$ &
$2.379$ & $-$\\
&  & Ce(4) & $2.251$ & $2.309$ & $2.361$ & $2.544$ & $2.267$ & $2.279$ &
$2.356$ & $-$\\
&  & Ce(5) & $2.429$ & $2.254$ & $2.268$ & $2.342$ & $2.240$ & $2.358$ &
$2.281$ & $-$\\
&  & Ce(6) & $2.214$ & $2.280$ & $2.227$ & $2.315$ & $2.306$ & $2.274$ &
$2.429$ & $-$\\
Ce$_{6}$O$_{11}$ & \onlinecite{Sorensen1976,Villars1991,Zhang1996} & Ce(1) &
$2.180$ & $2.241$ & $2.245$ & $2.490$ & $2.560$ & $2.618$ & $-$ & $-$\\
&  & Ce(2) & $2.189$ & $2.230$ & $2.236$ & $2.252$ & $2.516$ & $2.669$ &
$2.834$ & $-$\\
&  & Ce(3) & $2.260$ & $2.324$ & $2.377$ & $2.441$ & $2.593$ & $2.603$ &
$2.698$ & $-$\\
&  & Ce(4) & $2.064$ & $2.143$ & $2.147$ & $2.189$ & $2.670$ &  & $-$ & $-$\\
&  & Ce(5) & $2.235$ & $2.285$ & $2.347$ & $2.482$ & $2.723$ & $2.830$ & $-$ &
$-$\\
&  & Ce(6) & $2.180$ & $2.241$ & $2.245$ & $2.490$ & $2.590$ & $2.618$ & $-$ &
$-$\\
CeO$_{2}$ & \onlinecite{Villars1991,Taylor1984,Sorensen1976} & Ce & $2.434$ &
$2.434$ & $2.434$ & $2.434$ & $2.434$ & $2.434$ & $2.434$ & $2.434$%
\\\hline\hline
\end{tabular*}
\label{lengths}%
\end{table*}%

\end{document}